\documentclass[letterpaper,floatfix,aps,pra,amsmath,amssymb,twocolumn,superscriptaddress,
showpacs,footinbib]{revtex4-1}

\usepackage{epsfig}
\usepackage{verbatim}
\usepackage{color}
\usepackage{epsf}
\usepackage{array}
\usepackage{amsmath}

\usepackage{graphicx}
\usepackage{hyperref}

\newcommand{\be}{\begin{equation}}
\newcommand{\ee}{\end{equation}}
\newcommand{\ba}{\begin{equation}}
\newcommand{\ea}{\end{equation}}
\newcommand{\bea}{\begin{eqnarray}}
\newcommand{\eea}{\end{eqnarray}}

\newcommand{\Do}{\Delta_0}

\newcommand{\e}{\epsilon}

\newcommand{\eref}[1]{Eq.~(\ref{#1})}
\newcommand{\esref}[1]{Eqs.~(\ref{#1})}
\newcommand{\rref}[1]{(\ref{#1})}

\newcommand{\ocite}[1]{Ref.~\cite{#1}}

\newcommand{\qp}{\mathrm{qp}}

\newcommand{\BCS}{\mathrm{BCS}}

\begin{document}

\title{Proximity effect in normal-metal quasiparticle traps}

\author{A. Hosseinkhani}
\affiliation{JARA Institute for Quantum Information (PGI-11), Forschungszentrum J\"ulich, 52425 J\"ulich, Germany}
\affiliation{JARA-Institute for Quantum Information, RWTH Aachen University, D-52056 Aachen, Germany}
\author{G. Catelani}
\affiliation{JARA Institute for Quantum Information (PGI-11), Forschungszentrum J\"ulich, 52425 J\"ulich, Germany}
\begin{abstract}
In many superconducting devices, including qubits, quasiparticle excitations are detrimental. A normal metal ($N$) in contact with a superconductor ($S$) can trap these excitations; therefore such a trap can potentially improve the devices performances. The two materials influence each other, a phenomenon known as proximity effect which has drawn attention since the '60s. Here we study whether this mutual influence places a limitation on the possible performance improvement in superconducting qubits. We first revisit the proximity effect in uniform $NS$ bilayers; despite the long history of this problem, we present novel findings for the density of states. We then extend our results to describe a non-uniform system in the vicinity of a trap edge. Using these results together with a phenomenological model for the suppression of the quasiparticle density due to the trap, we find in a transmon qubit an optimum trap-junction distance at which the qubit relaxation rate is minimized. This optimum distance,  of the order of 4 to 20 coherence lengths, originates from the competition between proximity effect and quasiparticle density suppression. 
We conclude that the harmful influence of the proximity effect can be avoided so long as the trap is farther away from the junction than this optimum.
\end{abstract}

\date{\today}


\maketitle

\section{Introduction}
\label{sec:intro}

Unwanted quasiparticle excitations can degrade the performance of superconducting devices. For example, they can be a limiting factor for superconducting detectors used in astronomy~\cite{Day}, and a source of errors in superconductor-based charge pumps used for metrology~\cite{PekolaRMP}. In superconducting qubits, quasiparticles tunneling across Josephson junctions interact with the qubit degree of freedom, leading to qubit relaxation. The quasiparticle-induced decay rate has been predicted \cite{GC_PRL106} and shown experimentally \cite{Paik} to linearly scale with the density of quasiparticles. In both resonators \cite{Visser} and qubits \cite{Riste} there is evidence for excess, non-equilibrium quasiparticles. While to our knowledge there is no agreed upon explanation for the origin of these excess quasiparticles at millikelvin temperatures, a number of different approaches have been explored in order to suppress them with the aim to improve device performance. These approaches include engineering the spatial profile of the superconducting gap to steer quasiparticles away from Josephson junctions \cite{Aumentado,Sun}, and cooling down a device in a magnetic field in order to generate vortices that trap quasiparticles in their cores~\cite{Nsanzineza,Wang,Taupin}. Here we consider a different way to capture quasiparticles using normal-metal traps, a proposal which has been implemented in single-electron turnstiles~\cite{Knowles},  hybrid multilayer coolers~\cite{Pekola2000,Rajauria}, superconducting resonators~\cite{Patel}, and qubits~\cite{Riwar}.

Normal-metal quasiparticle traps consist of a normal metal ($N$) in tunnel contact with a superconductor ($S$): quasiparticles that tunnel from the superconductor into the normal metal can either tunnel back, or lose energy to a level below the superconducting gap, in which case cannot return back to the superconductor. A model for the effective quasiparticle trapping rate has been presented theoretically and tested experimentally in Ref.~\cite{Riwar}. In the model, the superconductor is assumed to be described by BCS theory. However, when contacting $N$ and $S$ materials, Cooper pairs can ``leak'' into the normal part; this induces superconducting correlations inside the normal metal and suppresses the superconducting order parameter. These types of phenomena are known as proximity effect and its inverse, and they have been studied since the 1960~\cite{Gennes1,Gennes2,McMilan}. Works on the proximity effect have investigated, for example, quasi-one-dimensional $N$-$S$ systems~\cite{Gueron,belzig,Moussy}, $SNS$ junctions~\cite{kup,Zhou,Ivanov,Hammer,Sueur}, $NSN$ configurations~\cite{Kauppila}, and proximity between two different superconductors~\cite{Cherkez}. For our purposes, we note here that in an $NS$ bilayer a minigap in the single-particle density of states develops in both layers and a finite subgap density of states is induced in the superconducting layer~\cite{Fominov}. 
In this article we focus on the subgap states,
since it has been shown that their presence can increase the relaxation rate $1/T_1$ of a superconducting qubit~\cite{Leppakangas}. Our main result is that due to the competition between such a proximity effect-induced increase in the relaxation rate and the decrease of $1/T_1$ due to the trap's suppression of the quasiparticle density, there is an optimal position for the trap. If the trap is closer to a junction than this optimal position, the relaxation rate exponentially increases over a distance given by the coherence length; for a trap further away than the optimum, the decay rate slowly increases over the much longer ``trapping length'', which is determined by quasiparticle diffusion and the trap's effective trapping rate.



The paper is organized as follows: in Sec.~\ref{S2} we review qubit relaxation due to quasiparticles and summarize the model of normal-metal quasiparticle trapping. In Sec.~\ref{S3} we use the quasiclassical theory of superconductivity to study first a uniform normal-superconductor bilayer. Then (Sec.~\ref{S4}) we consider a non-uniform system which models quasiparticle traps; we present both numerical self-consistent solutions for the spatial variation of superconducting order parameter and single-particle density of states as well as approximate analytical expressions for the latter. In Sec. \ref{S5} we study the qubit decay rate taking into account the proximity effect.
We summarize our work in Sec.~\ref{S6}, while Appendices~\ref{app:usadal-non-unif} through \ref{App:Decay_rate} contain a number of derivations and mathematical details.

\section{Qubit relaxation due to quasiparticles}
\label{S2}

Superconducting qubits can store quantum information in a collective degree of freedom, the phase difference $\varphi$ across a Josephson junction. Such a phase difference induces a dissipationless supercurrent carried by Cooper pairs tunneling through the junction. However, the presence of quasiparticles (due to unpaired electrons) opens an unwanted decay channel: a quasiparticle that tunnels can exchange energy with the qubit and cause its decay. The effects of the qubit-quasiparticle interaction have been studied in detail theoretically in Ref.~\cite{GC_PRB11} -- we summarize here the relevant findings of that work. Focusing henceforth for simplicity on the case of a single-junction transmon, the quasiparticle contribution to the qubit decay rate,
$\Gamma_{10}$, can be written in the standard form of the product between a matrix element and a spectral density,
\be\label{eq:g10_1}
\Gamma_{10} = \left|\langle 1 |\sin\frac{\varphi}{2}|0\rangle \right|^2 S(\omega_{10})\, ,
\ee
where $|0\rangle$ ($|1\rangle$) denotes the ground (excited) state of the qubit, and $\omega_{10}$ is the qubit frequency. The excitation rate $\Gamma_{01}$ is obtained by replacing $\omega_{10} \to - \omega_{10}$. The matrix element can be expressed in terms of the transmon parameters as
\be\label{eq:trmel}
\left|\langle 1 |\sin\frac{\varphi}{2}|0\rangle \right|^2 = \frac{E_C}{\omega_{10}} \simeq \sqrt{\frac{E_C}{8E_J}}
\ee
with $E_C$ the charging energy and $E_J$ the Josephson energy; this expression is valid in the transmon regime $E_J \gg E_C$.

The spectral density takes a simple form under certain conditions that are often satisfied, namely: a hard gap $\Delta_0$ in the superconductor which is larger than the qubit frequency, $2\Delta_0 > \omega_{10}$; quasiparticles that are ``cold'', meaning that their typical energy $\delta E$ (or effective temperature) above the gap is small compared to qubit frequency, $\delta E \ll \omega_{10}$. Then we have
\be\label{eq:S_w_x_qp}
S(\omega_{10}) \simeq \frac{8E_J}{\pi} x_\qp \sqrt{\frac{2\Delta_0}{\omega_{10}}}
\ee
and $S(-\omega_{10}) \ll S(\omega_{10})$. In this expression, the prefactor proportional to $E_J$ accounts for the tunneling probability via the Ambegaokar-Baratoff relation $E_J= \Delta_0 g_T/8g_K$, where $g_T$ is the conductance of the junction and $g_K=e^2/2\pi$ is the conductance quantum. The central factor $x_\qp$ is the density of quasiparticles normalized by the Cooper pair density,
\be\label{eq:x_qp}
x_\qp = \frac{2}{\Delta_0} \int_{0} d\epsilon \, n(\epsilon) f(\epsilon)\,,
\ee
with $f(\epsilon)$ the quasiparticle distribution function and $n(\epsilon)$ the normalized density of states (DoS), which for a BCS superconductor is 
\be\label{eq:nBCS}
n(\epsilon) = n_\mathrm{BCS} (\epsilon) \equiv \frac{\epsilon}{\sqrt{\epsilon^2-\Delta_0^2}}\, .
\ee
The square root factor in \eref{eq:S_w_x_qp} accounts for the final density of states of quasiparticles, after tunneling and absorbing the qubit energy.

Equations \rref{eq:g10_1} and \rref{eq:S_w_x_qp} imply that suppressing the quasiparticle density near the junction can prolong the $T_1$ time, given by
\be\label{eq:T1_def}
\frac{1}{T_1} = \Gamma_{10} + \Gamma_{01} \, .
\ee
A way to suppress $x_\qp$ is by introducing normal-metal islands, in tunnel contact with the superconducting electrodes, which can trap quasiparticles. A model that accounts for the interplay between superconductor-normal island tunneling, energy relaxation in the normal metal, and diffusion in the superconductor was developed theoretically and tested experimentally in \ocite{Riwar}; in the model, the dynamics of the quasiparticle density is governed by a generalized diffusion equation
\be
\label{eq:diffusion}
\frac{\partial}{\partial t} x_\text{qp}=D_\text{qp}\nabla^2x_\text{qp}-a(\vec{r})\Gamma_\text{eff}x_\text{qp}
 + g \, ,
\ee
where $D_\qp$ is the quasiparticle diffusion constant, and the area function $a(\vec{r})$ equals 1 for coordinate $\vec{r}$ in the superconductor-normal metal contact region and 0 elsewhere. The quasiparticle generation rate $g$ phenomenologically accounts for all processes creating quasiparticles. The effective trapping rate $\Gamma_\text{eff}$ is determined by the balance between tunneling from superconductor to normal metal, the inverse escape process, and the energy relaxation of excitations in the normal metal. Experimentally, relaxation is the bottleneck limiting the effective trapping rate, see \ocite{Riwar}. More recent work has explored how to optimize trap performance by appropriately choosing the trap placement in a qubit~\cite{Hosseinkhani}.

All the works summarized so far rely on the hard-gap assumption. As mentioned in the introduction, and as we will explain in more detail in Sec.~\ref{S3}, due to the proximity effect the BCS peak in the density of states broadens and a finite subgap DoS is induced in the superconductor. Therefore, in this paper we want to relax the hard-gap assumption. As a first step, we consider how to generalize the expression for the qubit decay rate, \eref{eq:g10_1}; the appropriate generalization is presented in
\ocite{Leppakangas}, and for the case considered here of a single-junction transmon it amounts to a redefinition of the spectral density appearing in \eref{eq:g10_1}:
\be\label{eq:Sred}
S(\omega) = S_\mathrm{t} (\omega) + S_\mathrm{p} (\omega)\, ,
\ee
where we distinguish two contributions, $S_\mathrm{t}$ due to single quasiparticle tunneling and $S_\mathrm{p}$ originating from Cooper pair processes. In terms of the distribution function $f$ they are, for positive frequency $\omega>0$,
\bea\label{eq:S_t}
S_\mathrm{t}(\omega)&=&\int_{0}^{\infty}\!d\epsilon \, A(\epsilon,\epsilon+\omega) f(\epsilon)[1-f(\epsilon+\omega)], \\
S_\mathrm{p}(\omega)&=&\int_{0}^{\omega}\!d\epsilon \, \frac{1}{2} A(\epsilon,\omega-\epsilon) [1-f(\epsilon)][1-f(\omega-\epsilon)], \quad \quad  \label{eq:S_b}
\eea
with
\be
\label{eq:A}
A(\epsilon,\epsilon')=\frac{16E_J}{\pi\Delta_0}[n(\epsilon)n(\epsilon')+p(\epsilon)p(\epsilon')]\, .
\ee
The density of states $n(\epsilon)$ appearing in this expression does not necessarily take the BCS form. Both $n(\epsilon)$ and the pair amplitude $p(\epsilon)$ can be calculated within a Green's function approach, and in a normal/superconductor bilayer depend on parameters such as film thicknesses and interface resistance, as we explain next in Sec.~\ref{S3}. Here we point out that the combinations of $n$ and $p$ account for both the quasiparticle density of states and so-called coherence factors, while whether a process involves single quasiparticles or pairs is manifest in the combination of distribution functions: $f(1-f)$ for single quasiparticles, $(1-f)(1-f)$ or $ff$ for pair breaking or recombination processes, respectively. Finally, the spectral density $S$ at negative frequencies is obtained by replacing $f \to 1-f$ in \esref{eq:S_t} and \rref{eq:S_b}. For pair processes, this implies that  $S_\text{p}(\omega>0)$ accounts for pair breaking by qubit relaxation, while $S_\text{p}(\omega<0)$ for qubit excitation by quasiparticle recombination.


\section{Proximity effect in thin films}
\label{S3}

The goal of this section is to arrive at expressions for the functions $n$ and $p$ in \eref{eq:A} that take into account the proximity effect between the normal-metal trap and the qubit superconducting electrodes. These expression will then be used in Sec.~\ref{S5} to estimate the influence of the proximity effect on qubit lifetime. The calculations are based on the quasiclassical approach to superconductivity, which we briefly discuss in Appendix~\ref{app:usadal-non-unif} and is presented in more details in a number of reviews and textbooks~\cite{Rammer,Chandrasekhar,belzig2,Kopnin}. In this formalism, the properties of a superconductor can be encoded in the paring angle $\theta_S(\epsilon, r)$ which, for disordered superconductors, obeys an equation known as Usadel equation~\cite{Usadel}. Once $\theta_S$ is obtained, one can calculate quantities of interest such as the density of states and the pairing amplitude,
\bea
\label{eq:DOS}
n(\epsilon,r) & = & \mathrm{Re}[\cos \theta_S(\epsilon,r)], \\
p(\epsilon,r) & = & \mathrm{Im}[\sin \theta_S(\epsilon,r)]. \label{eq:p_def}
\eea
The Usadel equation must be supplemented by the self-consistent equation for the order parameter $\Delta(r)$, which assuming thermal equilibrium at temperature $T$ reads
\be
\label{eq:op_sc}
\Delta(T,r)=\frac{\nu_{0\mathrm{S}}\lambda}{2} \int_{-\omega_D}^{\omega_D}\!d\epsilon \, \tanh\left(\frac{\epsilon}{2T}\right) p(\epsilon,r) \, ,
\ee
where $\nu_{0S}$ is the density of states per spin at the Fermi energy in the normal state of the superconductor, and $\lambda$ is the effective coupling strength of the attractive electron-electron interaction responsible for superconductivity.
This formalism is applicable to non-uniform superconductors -- this enables us to study the behavior of the superconductor near the edge of a trap. However, we first consider proximity in a uniform bilayer.


\subsection{Uniform NS bilayers}
\label{sec:unifbi}

In a uniform bilayer a superconducting film of thickness $d_S$ is fully covered by a normal metal of thickness $d_N$, with both thicknesses smaller than the superconducting coherence length at zero temperature $\xi$. This implies that spatial variations across the films thicknesses can be neglected. Moreover,
because the system is uniform in the plane of the films, the pairing angle is independent of position and the Usadel equations take the form (see \ocite{Fominov} and Appendix~\ref{app:usadal-non-unif})
\bea
\label{eq:uni_usadelS}%
 i\epsilon\sin\theta_{S}(\epsilon) & + & \Delta(T)\cos\theta_S(\epsilon)  \\ &=& \frac{1}{\tau_S}\sin[\theta_{S}(\epsilon)-\theta_{N}(\epsilon)]\, , \nonumber \\
\label{eq:uni_usadelN}%
 i\epsilon\sin\theta_{N}(\epsilon) & = & \frac{1}{\tau_N}\sin[\theta_{N}(\epsilon)-\theta_{S}(\epsilon)]\, ,
\eea
where we have introduced a pairing angle $\theta_N$ for the normal layer and $\Delta(T)$ must be calculated self-consistently using \eref{eq:op_sc}.
The times $\tau_{i}= 2e^2\nu_i d_i R_{\mathrm{int}}A$ ($i=S,\,N$) account for the interface resistance times area product $R_\mathrm{int} A$ and the density of states at the Fermi level $\nu_i$ of the two films. Typically, $\tau_S \approx \tau_N$, and the dimensionless parameter $\tau_S \Delta$ can be used to characterize the strength of the coupling between the two layers. In the limit $\tau_S\Delta \to \infty$, to leading order we can neglect the right hand sides of \esref{eq:uni_usadelS} and \rref{eq:uni_usadelN}, so that the two layers would be decoupled; in this limit the solutions to the Usadel equations are
\bea\label{eq:thetaBCS}
\theta_S(\epsilon) &=& \theta_{BCS}(\epsilon)\equiv \arctan \frac{i\Delta}{\epsilon} \, , \\
\theta_N(\epsilon) &=& 0 \, .
\eea
For high interface resistance, such that $\tau_S\Delta \gg 1$ but finite, a weak-coupling regime is possible.  On the other hand, for a good contact between $N$ and $S$ [or sufficiently close to the critical temperature, where $\Delta(T) \to 0$] the coupling can be strong, $\tau_S \Delta \ll 1$. In this section we focus on the weak-coupling case $\tau_S\Delta \gg 1$ with $\tau_N \sim \tau_S$; some considerations on the strong-coupling one can be found in Appendix~\ref{App:Str_coup}.

\begin{figure}[t]
\begin{center}
\includegraphics[width=0.48\textwidth]{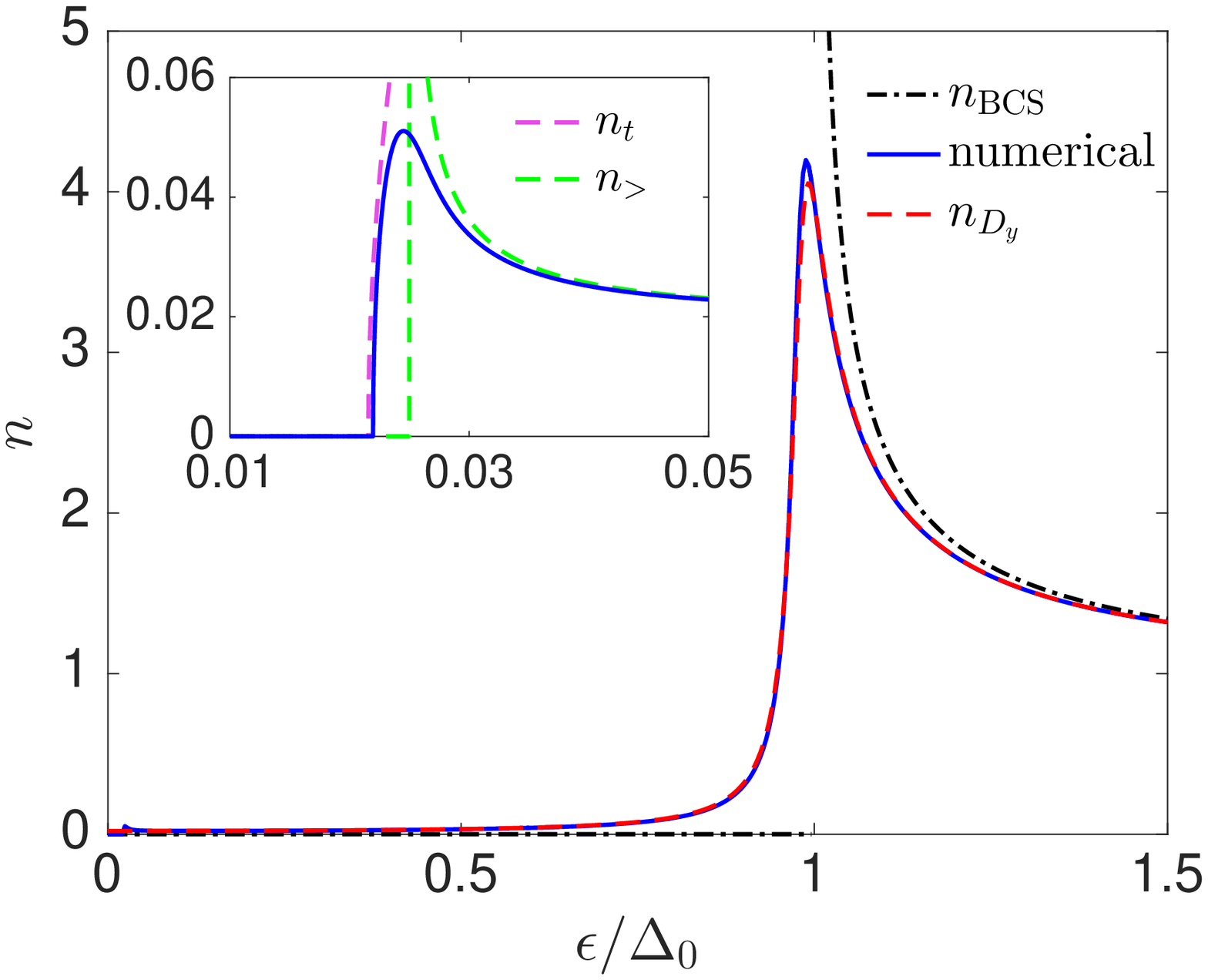}
\end{center}
\caption{(Color online) Density of states in a superconducting layer weakly coupled to a normal one, $\tau_S\Delta_0 = 50$. Solid lines (blue) are calculated by numerically solving the Usadel equations, \esref{eq:uni_usadelS} and \rref{eq:uni_usadelN}, and substituting the result into \eref{eq:DOS}. Dot-dashed line: BCS DoS, \eref{eq:nBCS}. Dashed lines: approximate analytical formulas: $n_{Dy}$ of \eref{eq:Dynes}, $n_t$ of \eref{eq:WeaCoup_DOS}, and $n_>$ of \eref{eq:nu_old}. The inset zooms to energies around the minigap.}  \label{fig:unif_weak_DOS_vs_E}
\end{figure}

In the normal film, the main consequence of the contact with the superconductor is the opening of a so-called minigap in its density of states. The minigap energy $E_g$ is always small compared to the gap in the bulk superconductor, and in the weak-coupling regime is given by
\be
E_g \simeq \frac{1}{\tau_N}\, ,
\ee
as already shown in the seminal work of McMillan~\cite{McMilan} and more recently rederived in \ocite{Fominov} within the quasiclassical formalism. In the superconductor, above the minigap a small but finite sub-gap density of states is induced of the form~\cite{McMilan,Fominov}
\be
\label{eq:nu_old}
n(\epsilon)\simeq n_>(\epsilon) \equiv \frac{1}{\tau_S\Delta} \mathrm{Re}\left[\frac{\epsilon}{\sqrt{\epsilon^2-1/\tau_N^2}}\right].
\ee
This expression is valid below the gap, $\epsilon \ll \Delta$, but it fails close to the minigap, as noted in \ocite{Fominov}. Indeed, as detailed in Appendix~\ref{app:unifweak} we find the validity condition
\be
\label{eq:Uni_validityCond}
\epsilon - \frac{1}{\tau_N} \gg \frac{1}{\tau_N} \mathrm{max}\left\{ \frac{1}{\left(\tau_S \Delta\right)^{2/3}}, \frac{1}{\tau_N \Delta} \right\}.
\ee
Moreover, the the position of the minigap is more accurately given by
\be\label{eq:minigap_WeaCoup}
\epsilon_g \simeq \frac{1}{\tau_N} \left[1 - \frac32\frac{1}{\left(\tau_S \Delta\right)^{2/3}}-\frac{1}{\tau_N\Delta}\right],
\ee
and just above it the density of states has a square root threshold behavior,
\be\label{eq:WeaCoup_DOS}
n(\epsilon) \simeq n_t(\epsilon)\equiv \frac{1}{\left(\tau_S\Delta\right)^{1/3}}\sqrt{\frac23 \tau_N \left(\epsilon - \epsilon_g \right)},
\ee
an expression valid for
\be
\tau_N (\epsilon-\epsilon_g) \ll (\tau_S\Delta)^{-2/3}.
\ee
In the inset of Fig.~\ref{fig:unif_weak_DOS_vs_E} we compare \esref{eq:nu_old} and \rref{eq:WeaCoup_DOS} to the density of states obtained by numerically solving the Usadel equations \rref{eq:uni_usadelS} and \rref{eq:uni_usadelN}.

We now turn our attention to energy well above the minigap, $\epsilon \gg \epsilon_g$. In this energy range a broadening of the BCS peak was qualitatively predicted~\cite{McMilan} and displayed in numerical calculations~\cite{Fominov}; to our knowledge, however, no analytical formula has been presented in the literature for the case of bilayers. Interestingly, we find that the density of states has the well-known form proposed by Dynes \textit{et al}.~\cite{Dynes} to fit tunneling measurements:
\be\label{eq:Dynes}
n(\epsilon) \simeq n_{Dy}(\epsilon) \equiv \mathrm{Re}\left[\frac{\epsilon + i/\tau_S}{\sqrt{(\epsilon + i/\tau_S)^2-\Delta^2}}\right].
\ee
A similar result was found for the case of a short $S$ wire between two $N$ leads~\cite{Kauppila}.
We obtain the above formula from the following approximate expression for the pairing angle
\be\label{eq:thetaDynes}
\theta_S (\epsilon) \simeq \theta_{Dy}(\epsilon) \equiv \arctan \frac{i\Delta}{\epsilon+i/\tau_S}
\ee
[deviations from these formulas can arise for $|\e/\Delta-1|\lesssim 1/(\tau_N\Delta)^2$ when $\sqrt{\tau_S \Delta} \gtrsim \tau_N \Delta$, see Appendix~\ref{app:unifweak}]. In the main panel of Fig.~\ref{fig:unif_weak_DOS_vs_E} we plot \eref{eq:Dynes} along with the result of a numerical calculation of the density of states.  Using $\theta_{Dy}$ of \eref{eq:thetaDynes} in the self-consistent equation \rref{eq:op_sc} we also recover McMillan result for the suppression of the zero-temperature order parameter,
\be\label{eq:D_NS}
\Delta_{NS} \simeq \Delta_0 \sqrt{1-\frac{2}{\tau_S\Delta_0}}\, ,
\ee
with $\Delta_0$ the bulk value of the order parameter.
Note that \eref{eq:nu_old} and \eref{eq:Dynes}
agree at leading order in the overlap region $\epsilon_g \ll \epsilon \ll \Delta$, as they both approximately take the constant value $1/\tau_S\Delta$ there; a crossover energy between the two expression can be identified with the geometric average $\sqrt{\Delta/\tau_N}$ between gap and minigap. This crossover energy is, for typical parameters, smaller than qubit frequency. Therefore, we can in general use the Dynes-like formulas as a starting point to evaluate the density of states in a non-homogenous system, which we consider next.

\subsection{Proximity effect near a trap edge}
\label{S4}

\begin{figure}[t]
\begin{center}
\includegraphics[width=0.41\textwidth]{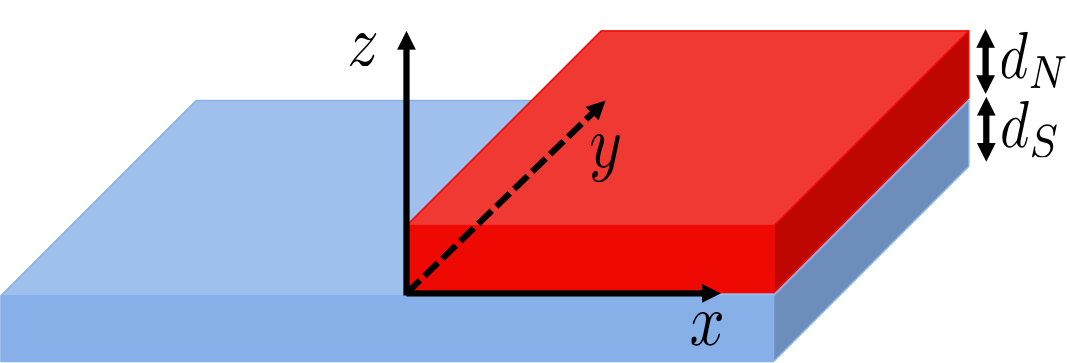}
\end{center}
\caption{(Color online) A non-uniform NS bilayer: a superconducting film of thickness $d_S$ (bottom) is partially covered by a normal metal layer (thickness $d_N$) occupying the region $x>0$. We use this system to model the vicinity of a normal-metal quasiparticle trap (see text).} \label{fig:systems}
\end{figure}

A normal-metal trap in general covers only part of a superconducting electrode [cf. \eref{eq:diffusion}], in order to limit losses in the normal metal that could otherwise shorten the qubit lifetime.
Typically, traps have lateral dimension of the order of $10~\mu$m or more~\cite{Riwar}, while the thicknesses $d_S$ and $d_N$ of superconducting and normal materials are in the range of tens of nanometers. These sizes should be compared to the coherence length $\xi$, which for disordered aluminum films typically used to fabricate qubits is of the order of 200~nm. Therefore both the normal and superconducting films are thin compared to $\xi$, while the lateral dimensions of the trap are much wider than $\xi$. We can therefore effectively model the system near the trap edge as being composed by a superconducting film occupying the whole $x$-$y$ plane and a normal metal in the half plane $x>0$, see Fig.~\ref{fig:systems}.

To study the proximity effect near such an edge, we must allow for spatially dependent paring angles. Due to translational symmetry in the $y$ direction, they are functions of coordinate $x$ only; they satisfy Usadel equations that generalize \esref{eq:uni_usadelS} and \rref{eq:uni_usadelN} by including a diffusion term
(see Appendix~\ref{app:usadal-non-unif} for the derivation):
\be \label{eq:non_uni-usadalS}\begin{split}
\frac{D_S}{2}\frac{\partial^2\theta_S(\epsilon,x)}{\partial x^2} &+ i\epsilon\sin\theta_{S}(\epsilon,x) + \Delta(x)\cos\theta_S(\epsilon,x) \\
& =  \frac{1}{\tau_S}\sin[\theta_S(\epsilon,x)-\theta_N(\epsilon,x)]H(x)\, ,
\end{split}\ee
where $D_S$ is the diffusion constant for electrons in the normal state of $S$ and $H(x)$ is the step function [$H(x)=1$ for $x>0$ and 0 otherwise], and for $x>0$
\be \begin{split} \label{eq:non_uni-usadalN}
\frac{D_N}{2}\frac{\partial^2\theta_N(\epsilon,x)}{\partial x^2} &+i\epsilon\sin\theta_{N}(\epsilon,x)  \\
& = \frac{1}{\tau_N}\sin[\theta_N(\epsilon,x)-\theta_S(\epsilon,x)],
\end{split}\ee
with $D_N$ the diffusion constant for electrons in $N$. As before, the superconducting order parameter $\Delta(x)$ is to be found self-consistently using \eref{eq:op_sc}. To avoid any confusion, we remind~\cite{Riwar} that while $D_\qp$ in \eref{eq:diffusion} is proportional to $D_S$ in \eref{eq:non_uni-usadalS}, the former takes into accounts phenomenologically the dependence on energy of the distribution function in the superconductor -- information that is lost in considering the density $x_\qp$ -- and this usually results in $D_\qp \ll D_S$.

\begin{figure}[t] 
\begin{center}
\includegraphics[width=0.48\textwidth]{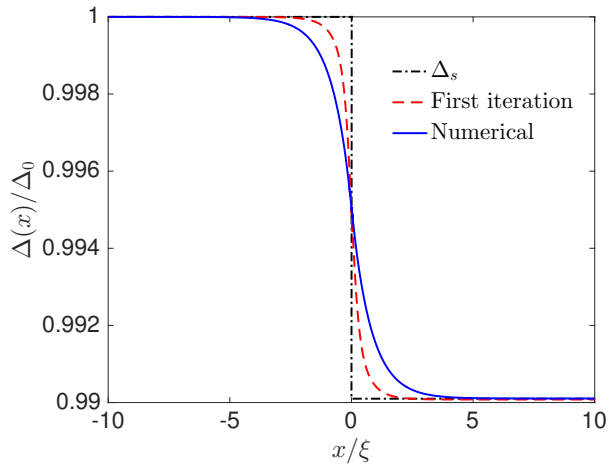}
\end{center}
\caption{(Color online) Solid line (blue): normalized order parameter $\Delta(x)/\Delta_0$ in the non-uniform NS bilayer depicted in Fig.~\ref{fig:systems} as a function of normalized distance $x/\xi$ from the trap edge. Dot-dashed line (black): non-self-consistent, step-like approximation, \eref{eq:OP_step}, which we use for analytical calculations (see text). Dashed line (red): ``first iteration'' obtained by substituting the pairing angles obtained in the step-like approximation, \esref{eq:Ana_theta_L} and \rref{eq:Ana_theta_R}, into \eref{eq:SemiAna_theta} and the latter into \eref{eq:op_sc}.}  \label{fig:non_uni_sc_op}
\end{figure}

In general, the system of Usadel plus self-consistent equations must be solved numerically. In Fig.~\ref{fig:non_uni_sc_op}, we plot with the solid line the self-consistent order paramter for such a solution, obtained following the procedure described in Appendix~\ref{app:num}. Far from the trap edge the solution must approach either the BCS one for $x\to-\infty$, or that for the uniform $NS$ bilayer for $x\to\infty$. In other words, indicating with $\theta_{Su}(\epsilon)$ the pairing angle in the $S$ component of a uniform $NS$ bilayer, the solution $\theta_S(\epsilon,x)$ for the non-uniform case interpolates between $\theta_{BCS}$ of \eref{eq:nBCS} and $\theta_{Su}$. Similarly, in the weak-coupling regime the order parameter $\Delta(x)$ interpolates between $\Delta_0$ and $\Delta_{NS}$ of \eref{eq:D_NS} as $x$ goes from $-\infty$ to $+\infty$; the difference between the two values of the order parameter is small, so we look for an approximate (not self-consistent) solution
to the Usadel equations \rref{eq:non_uni-usadalS} and \rref{eq:non_uni-usadalN} in which $\Delta(x)$ is assumed to take the form (see dot-dashed line in Fig.~\ref{fig:non_uni_sc_op})
\be\label{eq:OP_step}
  \Delta_s(x) = 
  \begin{cases}
    \Delta_{0}, &  x < 0\, , \\
    \Delta_{\mathrm{NS}}, &  x \geq 0\, .
  \end{cases}
\ee
Moreover, for energies large compared to the minigap, $\epsilon \gg \epsilon_g$, we can neglect $\theta_N$ in comparison to $\theta_S$ at leading order in $1/\tau_N\Delta_0 \ll 1$. Hence we can approximate $\sin[\theta_S-\theta_N] \approx \sin\theta_S$, and at this order \eref{eq:non_uni-usadalS} decouples from \eref{eq:non_uni-usadalN}. With these approximations, the solution for $\theta_S$ is (cf. Ref.~\cite{belzig})
\begin{align}
\label{eq:SemiAna_theta}
\theta_S(\epsilon,x) & = \theta_L(\epsilon,x)H(-x) + \theta_R(\epsilon,x)H(x)\, , \\
\theta_L(\epsilon,x) & =  \theta_{BCS}(\epsilon) \label{eq:SemiAna_theta_L} \\ 
& - 4\arctan\left\{e^{\frac{x}{\xi}\sqrt{2\alpha_1(\epsilon)}} \tan\left[\frac{\theta_{\BCS}(\epsilon)-\theta_0(\epsilon)}{4}\right]\right\}, \nonumber \\
\theta_R(\epsilon,x) & = \theta_{Su}(\epsilon) \label{eq:SemiAna_theta_R} \\ 
& - 4\arctan\left\{e^{-\frac{x}{\xi}\sqrt{2\alpha_2(\epsilon)}}\tan\left[\frac{\theta_{Su}(\epsilon)-\theta_0(\epsilon)}{4}\right]\right\}. \nonumber
\end{align}
Here, we define the coherence length as $\xi=\sqrt{D_S/\Delta_0}$, introduce the dimensionless functions
$\alpha_1(\epsilon)=\sqrt{\Delta_0^2-\epsilon^2}/\Delta_0$ and $\alpha_2(\epsilon)=\sqrt{\Delta_{NS}^2-(\epsilon+\frac{i}{\tau_S})^2}/\Delta_0$, and 
$\theta_0(\epsilon)$ is the (unknown) value of $\theta_S$ at the trap edge $x=0$. By construction this expression for $\theta_S$ is continuous at $x=0$, but it should also be continuously differentiable. Equating the left and right derivatives at the edge gives us a condition that implicitly defines $\theta_0$:
\be\label{eq:SemiAna_theta0}
\begin{split}
&\sqrt{\alpha_1(\epsilon)}\frac{\tan\left[\frac{\theta_{BCS}(\epsilon)-\theta_0(\epsilon)}{4}\right]}{1+\tan^2\left[\frac{\theta_{BCS}(\epsilon)-\theta_0(\epsilon)}{4}\right]} + \\
&\sqrt{\alpha_2(\epsilon)}\frac{\tan\left[\frac{\theta_{Su}(\epsilon)-\theta_0(\epsilon)}{4}\right]}{1+\tan^2\left[\frac{\theta_{Su}(\epsilon)-\theta_0(\epsilon)}{4}\right]} = 0\, .
\end{split}
\ee
In the weak-coupling regime we are considering, we have $\alpha_1 \simeq \alpha_2$ so long as $|\epsilon-\Delta_0| \gg 1/\tau_S$. Therefore, except in a narrow energy region near $\Delta_0$,
\eref{eq:SemiAna_theta0} has the approximate solution
\begin{align}
\label{eq:theta_0_fin}
\theta_0\simeq\frac{1}{2}(\theta_{BCS} + \theta_{Su}).
\end{align}
Finally, in the energy range where our approximations apply (energy above the minigap and not too close to $\Delta_0$), $\theta_{Su}$ is well approximated by $\theta_{Dy}$ of \eref{eq:thetaDynes}, which in the same energy range is close to $\theta_{BCS}$. We can therefore linearize \esref{eq:SemiAna_theta_L}  and \rref{eq:SemiAna_theta_R} to arrive at
\begin{align}
\label{eq:Ana_theta_L}
\theta_L(\epsilon,x) & \simeq \theta_{BCS}(\epsilon) - \frac{1}{2}e^{\frac{x}{\xi}\sqrt{2\alpha_1(\epsilon)}}\left[\theta_{BCS}(\epsilon)-\theta_{Su}(\epsilon)\right],\\
\label{eq:Ana_theta_R}
\theta_R(\epsilon,x) & \simeq \theta_{Su}(\epsilon) - \frac{1}{2}e^{-\frac{x}{\xi}\sqrt{2\alpha_2(\epsilon)}}\left[\theta_{Su}(\epsilon)-\theta_{BCS}(\epsilon)\right].
\end{align}

\begin{figure*}[!tb] 
\begin{center}
\includegraphics[width=\textwidth]{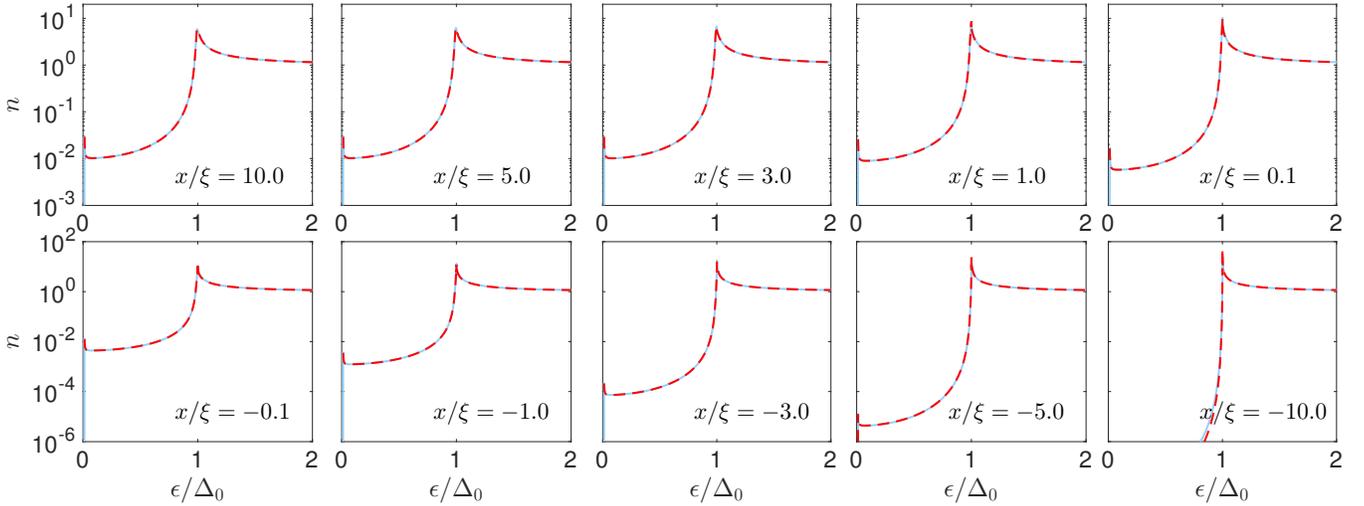}
\end{center}
\caption{(Color online) Density of states in the superconducting film of Fig.~\ref{fig:systems}, calculated for $\tau_S\Delta_0=100$ at various distances from the trap edge. The DoS takes the bilayer form far from the edge in the normal-metal covered region ($x/\xi=10$) and approaches the BCS expression exponentially fast as $x/\xi$ becomes more negative. We find excellent agreement between self-consistent numerical results (blue solid lines) and semi-analytical ones (red dashed lines); see text for details.}  \label{fig:SemiAn_VS_Num_DOS}
\end{figure*}

In Fig.~\ref{fig:SemiAn_VS_Num_DOS} we compare the density of states obtained from a self-consistent numerical solution of the Usadel equations (\ref{eq:non_uni-usadalS})-(\ref{eq:non_uni-usadalN}) to an approximate semi-analytic formula which we arrive at by substituting  Eqs.~(\ref{eq:Ana_theta_L})-(\ref{eq:Ana_theta_R}) [with $\theta_{Su}(\epsilon)$ found by numerically solving Eqs.~(\ref{eq:uni_usadelS})-(\ref{eq:uni_usadelN}) -- or equivalently Eq. (\ref{eq:X_full})] into \eref{eq:SemiAna_theta} and the latter into Eq.~(\ref{eq:DOS}). Our approximate formulas capture accurately the dependence of the DoS on the distance from the trap edge. While almost no deviations from the fully numerical calculation can be seen on the used scale, there are in fact differences in the region near the bulk gap, as expected -- see Appendix ~\ref{App:DOS_evolution}.


In the next section we will be interested in the spatial evolution of the normalized density of states and pair amplitude away from the normal-metal trap. In order to find analytical formulas for these quantities, we further approximate \eref{eq:Ana_theta_L} by using the Dynes expression, \eref{eq:thetaDynes}, for $\theta_{Su}$ and obtain for $x<0$ at leading order (see Appendix~\ref{App:DOS_evolution} for details)
\bea\label{eq:n_vs_x}
n(\epsilon,x) &\simeq& e^{-\sqrt{2}\frac{|x|}{\xi}\left(1-\frac{\e^2}{\Delta_0^2}\right)^{1/4}}\frac{1}{2}\frac{1}{\tau_S\Delta_0}\frac{\Delta_0^3}{\left(\Delta_0^2-\epsilon^2\right)^{3/2}}, \quad \\
\label{eq:p_vs_x}
p(\epsilon,x) &\simeq& e^{-\sqrt{2}\frac{|x|}{\xi}\left(1-\frac{\e^2}{\Delta_0^2}\right)^{1/4}}\frac{1}{2}\frac{1}{\tau_S\Delta_0}\frac{\e\Delta_0^2}{\left(\Delta_0^2-\epsilon^2\right)^{3/2}},
\eea
for $\Delta_0-\epsilon\gg 1/\tau_S$ and $\epsilon \gg \epsilon_g$, and
\bea
\label{eq:n_above}
n(\epsilon,x) &\simeq&  \frac{\epsilon}{\sqrt{\epsilon^2-\Delta_0^2}} -e^{-\frac{|x|}{\xi}\left(\frac{\e^2}{\Delta_0^2}-1\right)^{1/4}} \frac{1}{\sqrt{2}}\frac{1}{\tau_S\Do} \\ &\times&\frac{\Delta_0^3}{\left(\e^2-\Delta_0^2\right)^{3/2}}\cos\left[\frac{|x|}{\xi}\left(\frac{\e^2}{\Delta_0^2}-1\right)^{1/4}-\frac{\pi}{4}\right], \nonumber \\\label{eq:p_above}
p(\epsilon,x) &\simeq& \frac{\Delta_0}{\sqrt{\epsilon^2-\Delta_0^2}} -e^{-\frac{|x|}{\xi}\left(\frac{\e^2}{\Delta_0^2}-1\right)^{1/4}} \frac{1}{\sqrt{2}}\frac{1}{\tau_S\Do}\\
&\times&\frac{\e \Delta_0^2}{\left(\e^2-\Delta_0^2\right)^{3/2}}\cos\left[\frac{|x|}{\xi}\left(\frac{\e^2}{\Delta_0^2}-1\right)^{1/4}-\frac{\pi}{4}\right], \nonumber
\eea
for $\epsilon-\Delta_0 \gg 1/\tau_S$. Note that for energies above the gap the corrections to the BCS formulas are always small by construction. Moreover, both above and below the gap the corrections are small in $1/\tau_S\Delta_0$ and decay exponentially with distance over an energy-dependent length scale which is of the order of the coherence length $\xi$ away from the gap, but longer than $\xi$ close to the gap. We have now all the ingredients needed to estimate the quasiparticle-induced transition rates for a qubit with a trap, which is the focus of the next section.

\section{Qubit relaxation with a trap near the junction}
\label{S5}

As discussed in Sec.~\ref{S2}, the qubit decay rate due to quasiparticle tunneling is proportional to the spectral density $S(\omega)$, see \eref{eq:g10_1}. The spectral density is determined by the quasiparticle distribution function $f$, the density of states $n$, and the pair amplitude $p$, see \esref{eq:Sred}-\rref{eq:A}. We have shown in the previous Section that near a trap $n$ and $p$ become position-dependent. In the next subsection we study how this dependence affects the qubit decay rate, assuming that quasiparticles are everywhere in thermal equilibrium, so that the distribution function is uniform in space. This assumption is clearly not realistic, since it leads to an increase in the quasiparticle density approaching the trap, but it will enable us to show that the changes in the spectral density due to the proximity effect do not significantly harm the qubit if the trap is sufficiently far from the junction. In contrast, in Sec.~\ref{sec:effmu} we will account in a phenomenological way for the spatially dependent suppression of the quasiparticle density caused by the trap. In this more realistic scenario, we will find an optimal position for the trap, which balances between the density suppression and the enhancement of the subgap density of states, two effects that have opposite influence on the qubit relaxation rate.
Throughout this section, we assume that the qubit has reflection symmetry with respect to the junction, as in experiments~\cite{Riwar}; this means that when a trap is mentioned, it should be understood as two identical traps placed at the same distance from the junction. 

\subsection{Thermal equilibrium}
\label{sec:TherEqu}

The assumption of thermal equilibrium means that the distribution function has the Fermi-Dirac form,
\be\label{eq:feq}
f^{\text{eq}}(\epsilon) = \frac{1}{e^{\epsilon/T} + 1},
\ee
with $T$ the quasiparticle temperature. It then follows from \esref{eq:Sred}-\rref{eq:S_b} that the spectral density obeys the detailed balance relation
\be\label{eq:detbal}
S^{\text{eq}}(-\omega) = e^{-\omega/T}S^{\text{eq}}(\omega)\, .
\ee
We assume that the quasiparticles are ``cold'', $T\ll \omega_{10}$, and therefore we can neglect the qubit excitation rate in comparison with the decay rate, since $\Gamma_{01}^{\text{eq}}/\Gamma_{10}^{\text{eq}} = e^{-\omega_{10}/T} \ll 1$.

In presence of a trap, since $n$ and $p$ at the junction position depends on its distance $x$ from the trap, the quantity  $A$ defined in \eref{eq:A} is also a function of $x$ and so is the spectral function. An approximate expression for $S^\text{eq}(\omega,x)$ can be obtained in the relevant regime $\epsilon_g \ll T \ll \omega \ll \Delta_0$. In practice, since the minigap energy $\epsilon_g$ is much smaller than temperature $T$, we can set the former to zero. Then for the quasiparticle tunneling part of the spectral density, we can identify three contributions (see Appendix~\ref{App:Decay_rate} for details on the derivation of the expressions discussed here):
\be\label{eq:St_eq}
S^\text{eq}_\text{t}(\omega,x) = S_{aa}^\text{eq}(\omega,x) + S_{ba}^\text{eq}(\omega,x)+ S_{bb}^\text{eq}(\omega,x).
\ee
The first contribution accounts for transitions in which the initial quasiparticle energy is above the gap -- then the final energy is also above the gap; this term is approximately independent of position [cf. \eref{eq:S_w_x_qp}],
\be\label{eq:Saa}
S_{aa}^\text{eq}(\omega,x) \simeq \frac{8E_J}{\pi}x_\qp^\text{eq}\sqrt{\frac{2\Delta_0}{\omega}},
\ee
where $x_\qp^\text{eq} = \sqrt{2\pi T/\Delta_0} e^{-\Delta_0/T}$ coincides with the equilibrium value of the quasiparticle density in the absence of the trap.
A spatial dependence in principle arises from the corrections terms in \esref{eq:n_above} and \rref{eq:p_above}, but their contributions can be neglected in comparison with the other terms in $S^\text{eq}_\text{t}$ which we now discuss.

The second term in the right hand side of \eref{eq:St_eq} originates from transitions in which a quasiparticle initially below the gap absorbs the qubit energy and is excited above the gap energy:
\be\label{eq:Sba}
S_{ba}^\text{eq}(\omega,x) \simeq \frac{8E_J}{\pi}x_\qp^\text{eq}\sqrt{\frac{2\Delta_0}{\omega}}\frac{1}{2\tau_S\Delta_0} e^{-\sqrt{2}\frac{x}{\xi}\left(\frac{2\omega}{\Delta_0}\right)^\frac14}\frac{\Delta_0}{\omega}e^{\omega/T}.
\ee
The small factor $1/\tau_S\Delta_0$ and that exponentially decaying with distance account for the smallness of the initial density of states. In contrast, the final factor is large because the initial occupation probability is exponentially larger at lower energies. Thus at sufficiently low temperature this term can become larger than $S_{aa}$ of \eref{eq:Saa}. 

The last term in \eref{eq:St_eq} arises from transitions with both initial and final quasiparticle energy below the gap,
\be\label{eq:Sbb}
S_{bb}^\text{eq}(\omega,x) \simeq \frac{8E_J}{\pi}\frac{1}{\left(\tau_S\Delta_0\right)^2}e^{-2\sqrt{2}\frac{x}{\xi}}2\ln (2) \frac{T}{\Delta_0}\, .
\ee
Here the small factor $1/\tau_S\Delta_0$ is squared, and the exponential decay with distance is faster than in $S_{ba}^\text{eq}$ of \eref{eq:Sba}, because both initial and final density of states are small. However, the temperature dependence is much weaker: $S_{bb}^\text{eq}$ vanishes linearly with $T$ rather than exponentially, as $S_{ba}^\text{eq}$ does. Therefore, despite the small prefactors, this term can dominate at low temperatures.

\begin{figure}[t!] 
\includegraphics[width=0.48\textwidth]{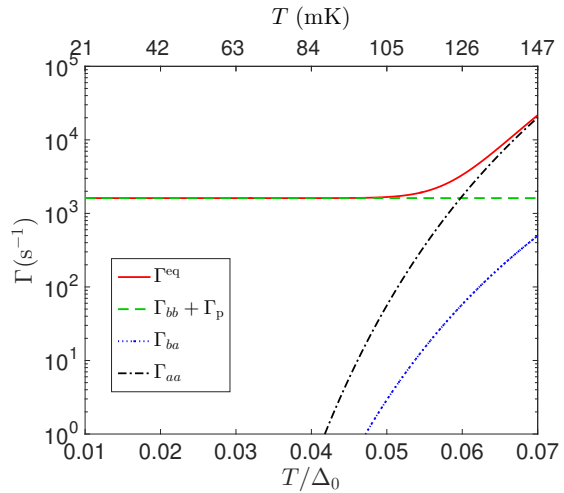}
\caption{(Color online) Qubit relaxation rate as a function of temperature. We assumed typical transmon parameters for the qubit (see \textit{e.g.} Ref.~\cite{Paik}): $\Delta_0=46\,$GHz, $\omega_{10}=6\,$GHz, $E_J=16\,$GHz and $E_C=290\,$MHz; and weak proximity effect, $\tau_S\Delta_0=10^3$. The solid line (red) shows the total relaxation rate, while the other lines show the contributions from the different processes discussed in the text.}  \label{fig:rates_TherEqui}
\end{figure}

In addition to the single quasiparticle tunneling, pair events can take place. In particular, since the density of states is finite (albeit small) down to the minigap energy $\epsilon_g$, so long as $\omega_{10}> 2\epsilon_g$ a pair breaking process is possible, in which the qubit relaxes by breaking a Cooper pair and exciting two quasiparticles above the minigap (but well below the gap). From \eref{eq:S_b} the spectral density for such a process is (see Appendix~\ref{App:Decay_rate})
\be\label{eq:Sp_eq}
S_\text{p}^\text{eq} (\omega,x)= \frac{8E_J}{\pi} \frac{1}{\left(\tau_S\Delta_0\right)^2}e^{-2\sqrt{2}\frac{x}{\xi}}\left[\frac{\omega}{\Delta_0} - 
2\ln(2) \frac{T}{\Delta_0}\right].
\ee
The spectral density does not vanish even at $T=0$, as there is no need for thermally excited quasiparticles to be present; in fact, the spectral density decreases linearly with increasing $T$ because the increased occupation of the final states suppresses this process. Interestingly, this linear in temperature term cancels with $S_{bb}$, \eref{eq:Sbb}; moreover, while $S_{ba}$ in \eref{eq:Sba} can be dominant in the limits of sufficiently small temperature and large distance, its contribution to $S_\text{t}^\text{eq}$ is negligible in the parameter range we are interested in (cf. Fig.~\ref{fig:rates_TherEqui}), so that we have approximately
\be\label{eq:Sthapp}
S^\text{eq}(\omega,x) \approx \frac{8E_J}{\pi} \left[x_\qp^\text{eq} \sqrt{\frac{2\Delta_0}{\omega}}+\frac{1}{\left(\tau_S\Delta_0\right)^2}\frac{\omega}{\Delta_0}e^{-2\sqrt{2}\frac{x}{\xi}}\right].
\ee

Assuming that the trap is next to the junction, $x=0$, in Fig.~\ref{fig:rates_TherEqui} we plot, as a function of temperature, the qubit decay rate $\Gamma^\mathrm{eq}$ obtained by substituting \esref{eq:trmel} and \rref{eq:Sthapp} into \eref{eq:g10_1}, as well as the contributions from the processes discussed above (above gap to above gap, $aa$; below gap to above gap, $ba$; and the sum of below gap to below gap, $bb$, with pair, $\text{p}$). At ``high'' temperature, above about 120~mK but still below the qubit frequency, the dominant contribution comes from the position-independent $aa$ term. In contrast, at low temperature there is a temperature-independent plateau in $\Gamma_{10}$ originating from the sum of $bb$ and pair-breaking processes. This plateau shows that the trap can increase the decay rate exponentially in comparison with the no-trap rate, which coincides with the $aa$ term. However, the plateau is quickly suppressed by moving the trap away from the junction: for each coherence length increase in trap-junction distance, the plateau decreases by a factor $e^{2\sqrt{2}} \approx 17$. With the parameters of Fig.~\ref{fig:rates_TherEqui}, this means that for $x=4\xi$ the low-temperature decay rate would be of order $10^{-2}\,$Hz. Therefore, even though the trap adversely affects the qubit, the limitation imposed on the decay rate becomes quickly negligible by increasing the distance to the junction. The fact that the trap can only harm the qubit rather than improve its coherence is a consequence of the thermal equilibrium assumption. Next, we relax this assumption to find up to which point the trap can be beneficial.

\subsection{Suppressed quasiparticle density}
\label{sec:effmu}

A trap can be beneficial to a qubit primarily by suppressing the quasiparticle density at the junction~\cite{Hosseinkhani}, as discussed in Sec.~\ref{S2}. Within the phenomenological diffusion model of \eref{eq:diffusion}, the typical length scale over which such a suppression takes place is given by the trapping length $\lambda_\text{tr} = \sqrt{D_\qp/\Gamma_\text{eff}}$; this length scale is of order 100~$\mu$m~\cite{Riwar,Hosseinkhani}, much longer than the coherence length $\xi$. As for the strength of the proximity effect, based on the experimental parameters of Ref.~\cite{Riwar}, we estimate it to be $\tau_S\Delta_0 \sim 10^3$-$10^4$. The large separation of length scales together with the weakness of normal trap-superconductor coupling, $\tau_S\Delta_0 \gg 1$, make it possible to use \eref{eq:diffusion} to calculate the spatial profile of the density, while the modifications introduced by the proximity effect can be treated as corrections. Below we will consider a realistic device geometry when calculating the position-dependent density, but first we discuss how to incorporate such a non-equilibrium quasiparticle configuration into the evaluation of the qubit transition rates.

When we neglect the proximity effect, the density of states takes the BCS form, \eref{eq:nBCS}, and the quasiparticle density defined in \eref{eq:x_qp} can depend on position only through the distribution function $f$. Such dependence could arise, for example, due to a temperature profile. However, at low temperatures it is in general more appropriate to model non-equilibrium quasiparticles by introducing an effective chemical potential $\tilde\mu$~\cite{Owen} (which we measure from the Fermi energy). The reason is that recombination processes, which are needed for chemical equilibrium, are slower than the scattering processes responsible for thermalization~\cite{Scalapino}. Such a phenomenological non-equilibrium approach has been already considered in the qubit setting~\cite{Ansari}. Here to capture the spatial profile of the density we assume  the distribution function $f$ to have the form
\be\label{eq:fnoneq}
f(\epsilon)=\frac{1}{e^{(\epsilon-\tilde\mu)/\tilde{T}}+1}\, ,
\ee
where the effective chemical potential is a function of position,  $\tilde\mu=\tilde\mu(x)$, while the effective temperature $\tilde{T}$ is homogeneous and does not necessarily coincides with the phonon bath temperature. Indeed, typical quasiparticle densities in the absence of traps are in the range $x_\qp  \sim 10^{-7}$-$10^{-5}$, corresponding to effective temperatures (at $\tilde\mu=0$) from $\sim 145\,$mK to $\sim 200\,$mK, much higher than both the usual fridge temperature (10-20~mK) and the typical qubit temperature which is of order 35-60~mK~\cite{Vool,Jin}, as estimated from the excited state population. In the following we will present results for $\tilde{T}/\Delta_0$ in the range $0.01$ to $0.05$, corresponding to approximately 20~mK to 110~mK in aluminum; for a given effective temperature, the chemical potential can then be calculated by inverting \eref{eq:x_qp} [with $n(\epsilon)$ of \eref{eq:nBCS} and $f(\epsilon)$ of \eref{eq:fnoneq}]. So long as $e^{(\Delta_0 - \tilde\mu)/\tilde{T}}\gg 1$, the integration in \eref{eq:x_qp} gives approximately $x_\qp \simeq \sqrt{2\pi\tilde{T}/\Delta_0}e^{(\tilde\mu-\Delta_0)/\tilde{T}}$ and therefore we find
\be\label{eq:mu}
\tilde\mu(x) = \Delta_0 + \tilde{T} \ln\left[\sqrt{\frac{\Delta_0}{2\pi\tilde{T}}}x_\qp(x)\right].
\ee
The assumption made above gives a restriction on the range of allowed effective temperatures, $2\pi\tilde{T}/\Delta_0 \gg x_\qp^2$, which is however not relevant in practice since usually we have $x_\qp < 10^{-4}$~\cite{Wang} and $2\pi\tilde{T}/\Delta_0> 10^{-2}$ (since $\tilde{T}$ should be at least comparable to the fridge temperature). This restriction also implies $\tilde\mu < \Delta_0$. We will assume that in general the quasiparticle density $x_\qp(x)$ is larger than the thermal equilibrium value at temperature $\tilde{T}$, so that $\tilde\mu > 0$. Note that for a given $x_\qp$, $\tilde\mu$ is a decreasing function of $\tilde{T}$, while for a fixed $\tilde{T}$ it is an increasing function of $x_\qp$.

With the approach described above, given the quasiparticle effective temperature $\tilde{T}$ and the density profile $x_\qp(x)$, one can calculate the effective chemical potential $\tilde\mu(x)$ using \eref{eq:mu} and therefore obtain an expression for the non-equilibrium distribution function, \eref{eq:fnoneq}. Once the distribution function is known, we can evaluate the spectral density of \esref{eq:Sred}-\rref{eq:S_b}, which we denote hereinafter with $\tilde{S}$ to remind of its dependence on the  non-equilibrium parameters $\tilde{T}$ and $\tilde\mu$ (and hence on junction-trap distance; we drop in this section the variable $x$ as explicit argument of the spectral density for notational compactness). Similar to the thermal equilibrium case, in the single quasiparticle tunneling contribution $\tilde{S}_\text{t}$ we distinguish three terms: $\tilde{S}_t=\tilde{S}_{aa}+\tilde{S}_{ba}+\tilde{S}_{bb}$. For the first two terms on the right hand side we find, as discussed in Appendix~\ref{app:supp}, that they are proportional to $x_\qp$, as in thermal equilibrium:
\be
\label{eq:tSaa}
\tilde{S}_{aa}(\omega)\simeq \frac{8E_J}{\pi}x_\qp(x)\sqrt{\frac{2\Delta_0}{\omega}}
\ee
and
\be
\label{eq:tSba}\tilde{S}_{ba}(\omega) \simeq \frac{8E_J}{\pi}x_\qp(x)\sqrt{\frac{2\Delta_0}{\omega}}\frac{1}{2\tau_S\Delta_0} e^{-\sqrt{2}\frac{x}{\xi}\left(\frac{2\omega}{\Delta_0}\right)^\frac14}\frac{\Delta_0}{\omega}e^{\omega/\tilde{T}}.
\ee
Here we assume that $\omega>0$ and $\Delta_0 -\tilde\mu-\omega \gg \tilde{T}$; validity conditions for the approximations employed are discussed in more detail in Appendix~\ref{app:supp}.

For the term $\tilde{S}_{bb}$ we have different regimes depending on the ratio between $\omega$ and $\tilde\mu$:
\be\label{eq:tSbb}\begin{split}
&\tilde{S}_{bb} (\omega) \simeq \frac{8E_J}{\pi}\frac{1}{(\tau_S\Delta_0)^2}\times\\ &\left\{
\begin{array}{ll}
    e^{-2\sqrt{2}\frac{x}{\xi}} \frac{2\tilde{T}}{\Delta_0} \ln\left(\frac{1+e^{\tilde\mu/\tilde{T}}}{1+e^{(\tilde\mu-\omega)/\tilde{T}}}\right), & \tilde\mu \lesssim \omega ,  \\
    4e^{-2\sqrt{2}\frac{x}{\xi}\left(1-\frac{\tilde\mu^2}{\Delta_0^2}\right)^\frac14} \frac{\Delta_0^3\left(\Delta_0^2+\tilde\mu^2\right)}{\left(\Delta_0^2-\tilde\mu^2\right)^{3/2}\left(\Delta_0+\tilde\mu\right)^{3/2}} \times \\ \left(\frac{1}{\sqrt{\Delta_0-\mu-\tilde\omega}} - \frac{1}{\sqrt{\Delta_0-\tilde\mu}} \right), & \tilde\mu \gg \omega .
\end{array}
\right.
\end{split}\ee
For this term the similarity with thermal equilibrium is recovered only for $\tilde\mu \ll \tilde{T}$. This is the case also for the pair process contribution $\tilde{S}_\text{p}$:
\be\label{eq:tSp}\begin{split}
\tilde{S}_\text{p}(\omega) \simeq \frac{8E_J}{\pi} \frac{1}{\left(\tau_S\Delta_0\right)^2}e^{-2\sqrt{2}\frac{x}{\xi}} \frac{\tilde{T}}{\Delta_0} 
\frac{1}{1-e^{(2\tilde\mu-\omega)/\tilde{T}}}\times \\
\left[2\ln\frac{1+e^{(\omega-\tilde\mu)/\tilde{T}}}{1+e^{-\tilde\mu/\tilde{T}}}-\frac{\omega}{\tilde{T}}\right].
\end{split}\ee
However, a partial cancellation between $\tilde{S}_{bb}$ and $\tilde{S}_\text{p}$ takes place so long as $\omega-2\tilde\mu \gg \tilde{T}$, in which case we find
\be
\tilde{S}_{bb}(\omega) + \tilde{S}_\text{p}(\omega) \approx \frac{8E_J}{\pi}\frac{1}{(\tau_S\Delta_0)^2}\frac{\omega}{\Delta_0}e^{-2\sqrt{2}\frac{x}{\xi}}\, ,
\ee
as in thermal equilibrium [compare to the second term in \eref{eq:Sthapp}].

Turning now to the spectral density at negative frequencies, we note that a relation similar to \eref{eq:detbal},
\be
\tilde{S}_\text{t} (-\omega) = e^{-\omega/\tilde{T}} \tilde{S}_\text{t}(\omega)\, ,
\ee
follows from \esref{eq:S_t} and \rref{eq:fnoneq}. Since we consider $\omega \gg \tilde{T}$, we can neglect the qubit excitation due to single quasiparticle tunneling. In contrast, for pair processes we find, from \eref{eq:S_b},
\be\label{eq:tSpneg}
\tilde{S}_\text{p} (-\omega) = e^{(2\tilde\mu-\omega)/\tilde{T}} \tilde{S}_\text{p}(\omega)\, .
\ee
Therefore the rate of qubit excitation induced by quasiparticle recombination can become exponentially larger
than qubit relaxation by Cooper pair breaking if $2\tilde\mu-\omega \gg \tilde{T}$. We next apply these results to a model of an actual qubit.

\subsubsection{An example}

\begin{figure}[t] 
\begin{center}
\includegraphics[width=0.48\textwidth]{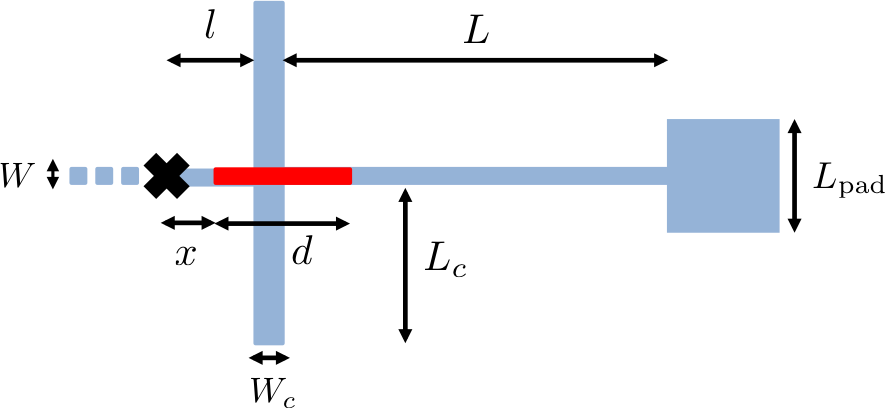}
\end{center}
\caption{(Color online) Diagram for the right half of the transmon qubit considered here. Except for the position of the trap, it is the same design studied in Refs.~\cite{Riwar,Hosseinkhani}.}  \label{fig:trap_transmon}
\end{figure}

As a concrete example, we consider the qubit geometry depicted in Fig.~\ref{fig:trap_transmon}, where a trap with length $d$ is placed a distance $x$ away from the Josephson junction. Similar geometries have been used experimentally to measure quasiparticle recombination, trapping by vortices~\cite{Wang} and by normal-metal traps~\cite{Riwar}, and theoretically to devise how to optimize trap performance~\cite{Hosseinkhani}. Here the only difference is that we allow the long trap, $d>l$, to be close to the junction, $0\le x \le l$, so that the role of the proximity effect can be evaluated. 

To find the quasiparticle density $x_\qp(x)$ at the Josephson junction, we proceed as in \ocite{Hosseinkhani} and treat each segment of the device (except the pad with side $L_\text{pad}$) as one-dimensional. Since we are interested in the steady-state density, we set $\partial x_\qp/\partial t=0$ in \eref{eq:diffusion}. Solving that equation for each segment  of the device, and requiring continuity of the density and current conservation at the points where different segments meet, we find
\begin{align}
\label{eq:x_qp_diffusion}
x_\qp(x)  = & \frac{g}{\Gamma_{\mathrm{eff}}}\bigg\{ 1 + \frac{1}{\sinh (d/\lambda_\mathrm{tr})}\bigg[ \frac{A_R}{W\lambda_\mathrm{tr}} + \cosh \left(\frac{d}{\lambda_{tr}}\right)\frac{x}{\lambda_\mathrm{tr}}  \nonumber \\ &+  \cosh\left(\frac{x+d-l}{\lambda_\mathrm{tr}}\right)\frac{A_c}{W\lambda_\mathrm{tr}} \bigg] + \frac{1}{2}\left(\frac{x}{\lambda_{\mathrm{tr}}}\right)^2 \bigg\} .
\end{align}
Here $A_{R}=W(L+l-x-d) + L_\mathrm{pad}^2$ is the uncovered area to the right of the trap and  $A_c=2W_cL_c$ is the area of gap capacitor. Equation~\rref{eq:x_qp_diffusion} makes it clear that the closer the trap is to the junction, the more the quasiparticle density is suppressed, and that significant changes in the density take place over the length scale given by $\lambda_\mathrm{tr}$.

\begin{figure}[!tb]
\begin{center}
\includegraphics[width=0.48\textwidth]{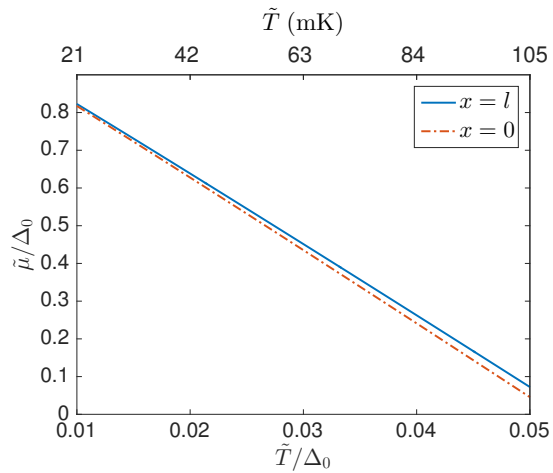}
\end{center}
\caption{(Color online) The normalized effective chemical potential $\tilde\mu/\Delta_0$, see \eref{eq:mu}, as function of the normalized effective temperature $\tilde{T}/\Delta_0$ for two positions of the trap: $x=0$ (dashed line) and $x=l$ (solid). The upper effective temperature scale is given for aluminum.}  \label{fig:mu_vs_x}
\end{figure}

We can now proceed as outlined above: namely, we first calculate the effective chemical potential $\tilde\mu$ for different effective temperatures $\tilde{T}$. In Fig.~\ref{fig:mu_vs_x} we show the results of such calculations for two positions of the trap, $x=0$ and $x=l$; hereinafter we use the same realistic parameters for the qubit ($L=1\,$mm, $l=60\,\mu$m, $W=12\,\mu$m, $L_c=200\,\mu$m, $W_c=20\,\mu$m, $L_\text{pad}=80\,\mu$m) and for the trap ($d=234\,\mu$m, $\lambda_\text{tr}=86.2\,\mu$m) as in Ref.~\cite{Hosseinkhani}, cf. Refs.~\cite{Wang,Riwar}. We also use the experimentally determined values $g=10^{-4}\,$Hz~\cite{Wang} and $\Gamma_\mathrm{eff} = 2.42\times10^{5}\,$Hz~\cite{Riwar}. As expected, the effective chemical potential decreases with increasing effective temperature, and is larger when the trap is further away. 

\begin{figure}[!tb] 
\begin{center}
\includegraphics[width=0.48\textwidth]{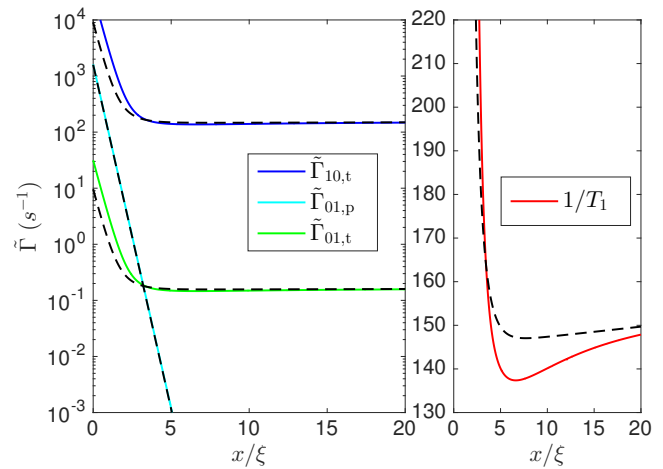}
\end{center}
\caption{(Color online) Left: tunneling (t) and pair (p) contributions to the qubit decay rate $1/T_1$ as a function of trap-junction distance $x$ (the pair decay rate $\tilde\Gamma_{10,\mathrm{p}}$ is too small to be visible). The effective temperature is $\tilde{T}/\Delta_0 = 0.019$ (approximately 40~mK in Al), other parameters are as in Fig.~\ref{fig:rates_TherEqui}.
The solid lines have been obtained by numerically calculating the integrals determining the spectral functions, while the dashed lines are our approximate analytical findings (see text for more details). Right: Total qubit decay rate, showing a minimum for a distance of a few coherence lengths.}  \label{fig:decay_NonEqu}
\end{figure}

Next, we calculate the qubit transition rates using the thus found chemical potential. We perform the calculation in two ways: we substitute the chemical potential into \eref{eq:fnoneq} for the distribution function and the latter into the definitions of the spectral functions, \esref{eq:S_t} and \rref{eq:S_b}; in those equations, we use the semi-analytic results for the pairing angle to determine the density of states and the pair amplitude, see \eref{eq:Ana_theta_L} and the text below \eref{eq:Ana_theta_R}, and perform the final integration over energy numerically. In a second approach, we use our approximate analytical formulas for the spectral functions, see \esref{eq:tSaa} to \rref{eq:tSpneg} (in deriving these formulas additional approximations were introduced, so the results are less accurate). The rates so obtained are shown in Fig.~\ref{fig:decay_NonEqu} for an effective temperature $\tilde{T}/\Delta_0 = 0.019$ ($\sim 40\,$mK in Al). In the left panel we distinguish the contributions to $1/T_1$ due to tunneling-induced relaxation ($\tilde\Gamma_{10,\mathrm{t}}$) and excitation ($\tilde\Gamma_{01,\mathrm{t}}$)  and pair process excitation ($\tilde\Gamma_{01,\mathrm{p}}$); the pair process relaxation rate is much smaller than the excitation rate [cf. \eref{eq:tSpneg}] and not visible on this scale. In the right panel we plot the total rate $1/T_1$, which is dominated by the tunneling-induced relaxation; the total rate is a non-monotonic function of trap-junction distance $x$, due to the competition between processes with initial quasiparticle energy below the gap, whose contributions to the rate decay exponentially with distance over a length scale of the order of the coherence length [see \esref{eq:tSba} and \rref{eq:tSbb}], and processes with above-gap initial energy, with contribution slowly increasing with $x_\qp$ over the much longer length scale $\lambda_\text{tr}$. The minimum in this curve thus gives the optimal position $x_o$ for the trap: for $x>x_o$, the density slowly increases, so one is not taking full advantage of the trap, but at $x<x_o$ the subgap density of states quickly increases and negates the benefit of further density suppression. Therefore we conclude that placing a trap at a distance $x_o<x<\lambda_\text{tr}$ represents the best choice.

\begin{figure}[t!]
\begin{center}
\includegraphics[width=0.48\textwidth]{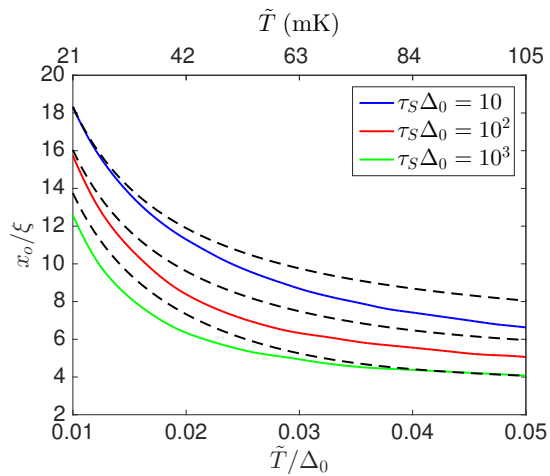}
\end{center}
\caption{(Color online) Normalized optimal trap-junction distance $x_o/\xi$ as a function of normalized effective temperature $\tilde{T}/\Delta_0$ (the upper scale gives the corresponding temperature for aluminum). Solid (dashed) lines are obtained by numerically finding the position of the minimum in curves such as the solid (dashed) one in the right panel of Fig.~\ref{fig:decay_NonEqu}.}  \label{fig:X_optm_vs_T}
\end{figure}

In Fig.~\ref{fig:X_optm_vs_T} we further explore the dependence of the optimal position $x_o$ on parameters such as the effective temperature $\tilde{T}$ and the strength of the proximity effect $\tau_S\Delta_0$ (we remind that the larger this parameter, the weaker the proximity effect). 
Clearly, the stronger the proximity effect, the further a trap should be placed. As the effective temperature increases, on the other hand, the optimal position decreases: as already noted before, for a given density the higher the effective temperature, the smaller the effective chemical potential, and this reduces the importance of the subgap states, since their occupation decreases. Based on this figure, we conclude that when the distance is over 20 coherence lengths the proximity effect can be safely neglected; in aluminum such a distance is of the order of a few microns, which is still much less than the trapping length.


\section{Summary}
\label{S6}

In this work we investigate the proximity effect between a normal-metal quasiparticle trap and the superconducting electrode of a qubit. On one hand, a trap can prolong the relaxation time of the qubit by suppressing the quasiparticle density. On the other hand, the proximity effect induces subgap states which can shorten the relaxation time. To quantify the competition between these two phenomena, we start by considering a uniform superconductor-normal metal bilayer; at relevant energies, the density of states takes the Dynes form, \eref{eq:Dynes}, with the broadening determined by interface resistance, superconducting film thickness, and its density of states at the Fermi energy. We then study how such broadening decays away from a trap edge, see \esref{eq:n_vs_x} to \rref{eq:p_above}. With these results, we can evaluate the qubit decay rate as function of the distance between trap and junction; we take into account the suppression of the quasiparticle density by introducing a distribution function which depends on two parameters, an effective temperature and a distance-dependent effective chemical potential, cf. \eref{eq:fnoneq}. Within this approach, we find that the competition between proximity effect and density suppression leads to an optimal placement for the trap, see Figs.~\ref{fig:decay_NonEqu} and \ref{fig:X_optm_vs_T}. The qubit relaxation rate exponentially increase for a trap closer to the junction than this optimum over a length scale of the order of the coherence length, while the increase of the rate when moving the trap farther away is much slower and over the much longer trapping length. Therefore, a trap should be placed at least as far from the junction as the optimum position, but no significant penalty is paid for distances up to the trapping length.

While we focused here on a transmon qubit, our findings may prove useful in designing traps for other systems as well. For example, quasiparticle poisoning could be a significant hurdle for nanowire-based realization of Majorana qubits~\cite{loss}; our findings indicate that a normal-metal trap placed close to the ends of the nanowire could be detrimental, as the small minigap is not sufficient to protect zero-energy states from being thermally excited into the subgap states induced by the trap. Finally, our results on the proximity effect could also help interpreting tunneling density of state experiments such as those reported in Refs.~\cite{Moussy,Rod}

\acknowledgments

We gratefully acknowledge useful discussions with M. Ansari, W. Belzig, L. Glazman, I. Khaymovich, H. Pothier, and R. Riwar.
This work was supported in part by the European Union under Research Executive Agency grant agreement No. CIG-618258.

\appendix
\section{Usadel equations for NS bilayers}
\label{app:usadal-non-unif}

In this Appendix we briefly derive the Usadel equation for uniform and non-uniform bilayers which are used in Sec.~\ref{S3}. Our starting point is 
the quasiclassical theory of superconductivity, a well-established approach to describe conventional superconductors~\cite{Rammer,Chandrasekhar,belzig2,Kopnin}. For our purposes, we need to consider the retarded Green's function $\hat{R}(r,\e)$ which is a $2\times2$ matrix in electron-hole space and satisfies the normalization condition  $\hat{R}^2(r,\e) = 1 $ . In the ``dirty'' limit in which the mean free path is smaller than the superconducting coherence length, and neglecting spin-flip and inelastic scatterings, the equation satisfied by the retarded Green's function is the so-called Usadel equation ~\cite{Fominov,Usadel}:
\begin{eqnarray}
\label{eq:G_usadal_equation}
\hbar D_i\nabla(\hat{R}_i\nabla\hat{R}_i)+i[\e\hat{\sigma}_z + \hat{\Delta},\hat{R}_i]=0\;  ,
\end{eqnarray}
where index $i=S,\,N$ is used to denote superconducting or normal material, $D_i$ is the (normal-state) diffusion constant, $\hat{\sigma}_z$ is the Pauli matrix in electron-hole space, and the matrix $\hat{\Delta}(r)$ has the form
\begin{eqnarray}
\label{eq:op-matrix}
\hat{\Delta}=
 \begin{pmatrix}
  0 & \Delta(r) \\
  -\Delta^*(r) & 0 
 \end{pmatrix} ,
\end{eqnarray}
where the superconducting order parameter $\Delta(r)$ is obtained from the self-consistent equation given in \eref{eq:op_sc} for a superconductor, and is zero inside a normal metal. Given the normalization condition for the Green's function, we can use the following angular parameterization 
\begin{align}
\label{eq:angular-param}
\hat{R_i}=
 \begin{pmatrix}
  \cos \theta_i(\e,r) & \ - i\sin \theta_i(\e,r) \\
  i\sin \theta_i(\e,r) & -\cos \theta_i(\e,r) 
 \end{pmatrix} \; 
\end{align}
which is appropriate in the absence of supercurrents (a phase factor would otherwise appear in the off-diagonal terms~\cite{Fominov}). Using this parameterization, the Usadel equation takes the form
\begin{align}  \label{eq:app:nonunif_usadel2}
&\dfrac{D_S}{2}\nabla^2\theta_S(\e,r) + i \e \sin\theta_S(\e,r)  \\
& \qquad + \Delta(r) \cos\theta_S(\e,r)=0, \nonumber \\
\label{eq:app:nonunif_usadel1}
&\dfrac{D_N}{2}\nabla^2\theta_N(\e,r) + i E \sin\theta_N(\e,r)=0 , 
\end{align}
in a superconductor and in a normal metal, respectively.

When the two materials are in contact, these equations must be supplemented by appropriate boundary conditions at the interface; these were derived by Kupriyanov and Lukichev~\cite{Lukichev} for contacts of low transparency (see also the end of Appendix~\ref{App:Str_coup}). Here we assume that a superconducting film of thickness $d_S$ (occuping the space between the planes at $z=0$ and $z=-d_S$, see Fig.~\ref{fig:systems}) is partially covered by a normal-metal film of thickness $d_N$ ($0<z<d_N$), with the NS contact in the $z=0$ plane; for coordinates in the contact region the boundary conditions read
\begin{align}
 \label{eq:app:bc}
& \sigma_N\frac{\partial\theta_N(\e,x,y,z)}{\partial z}\Big|_{z=0}=\sigma_S\frac{\partial\theta_S(\e,x,y,z)}{\partial z}\Big|_{z=0} \nonumber \\
& =\frac{1}{R_{\mathrm{int}}A}\sin[\theta_N(\e,x,y,0) - \theta_S(\e,x,y,0)],
\end{align}
where $A$ is the contact area, $R_\text{int}$ the interface resistance, and $\sigma_i=e^2\nu_i  D_i$ the normal-state conductivity ($\nu_i$ denotes the density of states at the Fermi level).  The normal derivatives of the angles $\theta_i$ at all other surfaces must vanish, thus ensuring current conservation.

We now take the thicknesses $d_i$ of the layers to be smaller than the zero-temperature superconducting coherence length $\xi$; this enables us to write the pairing angles as a series expansion in $z$. Using the vanishing of the normal derivatives at $z=-d_S$ and $z=d_N$ we find
\begin{align}\label{eq:app:angle2}
 \theta_S(\e,x,y,z) &= \overline{\theta}_S(\e,x,y) \nonumber \\
& +a(x,y)\alpha(\e,x,y)(\frac{z}{\xi} +\dfrac{z^2}{2\xi d_S})+...,
\end{align}
\begin{align} \label{eq:app:angle1}
 \theta_N(\e,x,y,z) &= \overline{\theta}_N(\e,x,y) \nonumber \\
 & + \beta(\e,x,y)(\frac{z}{\xi}-\dfrac{z^2}{2\xi d_N})+...,
\end{align}
where, as defined after \eref{eq:diffusion}, the area function $a(x,y)$ is 1 in the contact region and 0 otherwise (since the normal metal only partially covers the superconductor, $\theta_N$ is only defined in the contact region).
Substituting these two expressions into the Usadel equations (\ref{eq:app:nonunif_usadel2}) and (\ref{eq:app:nonunif_usadel1}) we obtain
\begin{align} 
\alpha(\e,x,y) & = -d_S\xi\bigg[\nabla^2\overline{\theta}_S(\e,x,y) 
 + \frac{2i\e}{D_S}\sin \overline{\theta}_S(\e,x,y) \nonumber \\ 
 & + \frac{2\Delta(x,y)}{D_S}\cos \overline{\theta}_S(\e,x,y) \bigg],\\
\beta(\e,x,y) & = d_N\xi\bigg[\nabla^2\overline{\theta}_N(\e,x,y)  
+\frac{2i\e}{D_N}\sin \overline{\theta}_N(\e,x,y)\bigg], \label{eq:app:beta}
\end{align}
where, from now on, the Laplacian acts in two-dimensional $x$-$y$ space.
Finally, we substitute Eqs.~(\ref{eq:app:angle1})-(\ref{eq:app:beta}) into the boundary conditions Eqs. (\ref{eq:app:bc}) and find:
\begin{align} \label{eq:app:usadal_3D_}
\frac{D_S}{2}&\nabla^2\overline{\theta}_S(\e,x,y)  +i\e\sin\overline{\theta}_S(\e,x,y) \nonumber \\
 & + \Delta(x,y)\cos\overline{\theta}_S(\e,x,y) \nonumber \\
 & =a(x,y)\frac{1}{\tau_S}\sin[\overline{\theta}_S(\e,x,y)-\overline{\theta}_N(\e,x,y)],
\end{align}
and
\be\label{eq:app:usadal_3D_1}
\begin{split} 
\frac{D_N}{2}\nabla^2\overline{\theta}_N(\e,x,y) +i\e\sin\overline{\theta}_N(\e,x,y) \\
 =\frac{1}{\tau_N}\sin[\overline{\theta}_N(\e,x,y)-\overline{\theta}_S(\e,x,y)],
\end{split}
\ee
where $\tau_{i}= 2e^2\nu_{i}d_{i}R_{\mathrm{int}}A$. In the case of a uniform bilayer, any spatial-dependency drops off and the above equations simplify to Eqs.~(\ref{eq:uni_usadelS}) and (\ref{eq:uni_usadelN}). In the non-uniform case depicted in Fig.~\ref{fig:systems}, the system is translationally invariant in the $y$ direction and Eqs.~(\ref{eq:app:usadal_3D_}) and (\ref{eq:app:usadal_3D_1}) take the form of Eqs.~(\ref{eq:non_uni-usadalS}) and (\ref{eq:non_uni-usadalN}).

\section{Proximity effect in uniform NS bilayers}

This Appendix has two parts: we first give some details of the calculation leading to the expressions presented in Sec.~\ref{sec:unifbi} for weakly-coupled uniform bilayers. In the second part we extend some of those results to stronger coupling.
Our firs step consist in the changes of variables $\theta_{i} = \pi/2 + i \chi_{i}$ and $\sinh\chi_S=X$ in Eqs.~(\ref{eq:uni_usadelS})-(\ref{eq:uni_usadelN}), leading to
\begin{align}
\label{eq:chi_N_1}
\sinh\chi_N&=\frac{\tau_N\e+X}{\sqrt{1+X^2}}\cosh\chi_N, \\
\label{eq:chi_N_2}
\cosh\chi_N&=-\frac{\tau_S}{\tau_N}\sqrt{1+X^2} + \frac{\tau_S\Delta}{\tau_N\e}X.
\end{align}
Squaring these equations and substituting the second one in the first gives
\begin{align} \label{eq:X_full}
\left(\frac{\tau_S\Delta}{\tau_N\e}\right)^2&\left[X-\frac{\e}{\Delta}\sqrt{1+X^2}\right]^2 \\
&\times\left[1 -2X\tau_N\e - \tau_N^2\e^2\right]=1+X^2. \nonumber
\end{align}
In what follow, we approximately solve this equation for $X$ as function of $\epsilon$; this enables us to find the normalized density of states, which in this notation is given by  $n(\epsilon)=\mathrm{Im}(X)$. 

\subsection{Weak-coupling limit}
\label{app:unifweak}

Let us consider the weak coupling limit $\tau_S \Delta, \, \tau_N\Delta \gg 1$. In the subgap region $\epsilon \ll \Delta$, the density of states is small, which suggest the assumption $X \ll 1$~\cite{Fominov}. If we further assume  
\be\label{furtherassump}
|2\tau_N\e X|\ll |1-\tau_N^2\e^2|\, ,
\ee
the solution to \eref{eq:X_full} is
\begin{align}
\label{eq:X_WeakCoup_aboveTher}
X=\frac{\e}{\tau_S\Delta\sqrt{1/\tau_N^2-\e^2}}+\frac{\e}{\Delta}.
\end{align}
This gives the minigap energy at $E_\mathrm{g}=1/\tau_N$ while the DoS above it is given by \eref{eq:nu_old}. As the energy approches $1/\tau_N$, however, $X$ becomes large, thus potentially violating the condition \rref{furtherassump}. Indeed, parameterizing the energy as $\tau_N \epsilon = 1+\kappa$ (with $0<\kappa \ll 1$), and using \eref{eq:X_WeakCoup_aboveTher}, \eref{furtherassump} takes the form
\begin{align}
\label{eq:kappa_condition}
\frac{1}{\tau_S\Delta\sqrt{2}\kappa^{3/2}}+\frac{1}{\tau_N\Delta\kappa} \ll 1.
\end{align}
Since both terms in the left hand side must be small, we arrive at \eref{eq:Uni_validityCond}. 

To study the DoS at energies below $1/\tau_N$, we must remove the  assumption \rref{furtherassump}, while still maintaining $X\ll 1$. Then, we can expand Eq.~(\ref{eq:X_full}) with respect to $X$, and keeping terms up to the cubic order we can write that equation in the form
\be
\mathcal{F}(X,\e) = 0
\ee
with
\begin{align} \label{eq:X_cubic}
 \mathcal{F}(X,\e) \equiv 2X^3 & +\tau_N\e\left(1-\frac{1}{\tau_N^2\e^2}-\frac{4}{\tau_N\Delta}\right)X^2 \nonumber\\ \nonumber
 & +2\tau_N\e\frac{\e}{\Delta}\left(\frac{1}{\tau_N^2\e^2}-1+\frac{1}{\tau_N\Delta}\right) X  \\
 & + \tau_N\e\left(\frac{1}{\tau_S^2\Delta^2} + \frac{\tau_N^2\e^2-1}{\tau_N^2\Delta^2}\right). 
\end{align}
Depending on its coefficient, a third order polynomial can have either three real roots, or one real and two complex conjugate roots. For the DoS not to vanish, we need $X$ to be complex, so the minigap is identified as the energy at which the type of roots changes from purely real -- this happens when the polynomial has a minimum so that the two real roots are degenerate, giving the condition 
$\frac{\partial\mathcal{F}(X,\e)}{\partial X}|_{\e=\e_g}=0$. Vanishing of the derivative requires
\begin{eqnarray}
\label{eq:X-critical-value}
X(\e_g)=-\frac{1}{3}\tau_N\e_g\left(1-\frac{1}{\tau_N^2\e^2_g}-\frac{1}{\tau_N\Delta}\right) 
\end{eqnarray}
[one can check that the second solution, $X(\e_g) = \e/\Delta$, leads to the unphysical result $\e_g=0$]. Substituting Eq.~(\ref{eq:X-critical-value}) into Eq.~(\ref{eq:X_cubic}) and solving for $\e_g$ at leading order in the small parameters $(\tau_N\Delta)^{-1}$ and $(\tau_S\Delta)^{-1}$, we arrive at \eref{eq:minigap_WeaCoup}.

To find the DoS above the minigap, we expand $X$ and $\tau_N\e$ around the minigap energy; we take $X=X(\e_\mathrm{g})+\delta X$ and $\e=\e_\mathrm{g}+\delta \e$ and expand Eq.~(\ref{eq:X_cubic}) up to first order in $\delta\e$ and second order in $\delta X$ (the lower orders vanish by construction):
\be
\label{eq:WeakCoup_expansion_general}
\mathcal{F}(X,\e)\simeq \frac{\partial\mathcal{F}}{\partial \e} (\delta \e)  + \frac{1}{2}\frac{\partial^2\mathcal{F}}{\partial X^2}(\delta X)^2  + \frac{\partial^2\mathcal{F}}{\partial \e\partial X}(\delta \e\delta X).
\ee
If the last term can be neglected, solving $\mathcal{F}(X,\e)=0$ for $\delta X$ in terms of $\delta \e$ clearly gives immediately a square root threshold behavior; the coefficients are given explicitly in \eref{eq:WeaCoup_DOS}. The applicability condition can be obtained \textit{e.g.} by requiring  $\frac{\partial^2\mathcal{F}}{\partial E\partial X}(\delta E\delta X) \ll \frac{\partial^2\mathcal{F}}{\partial^2 X}(\delta X)^2$, which gives
\begin{align}
\tau_N(\e-\e_\mathrm{g}) \ll (\tau_S\Delta)^{-2/3}.
\end{align}

We now consider energies much higher than the minigap, $\e \gg\e_\mathrm{g}$. In this case we assume 
\be\label{assumpDyn}
|1-2X\tau_N E|\ll |-\tau_N^2E^2|\, ,
\ee
and Eq.~(\ref{eq:X_full}) simplifies to $-\frac{\e}{\Delta}+\frac{X}{\sqrt{1+X^2}}=\frac{i}{\tau_S\Delta}$. Solving this equation for $X$ we arrive at the Dynes-like formula given in \eref{eq:Dynes}. Note that for $\e\gg\Delta$ the assumption \rref{assumpDyn} is always fulfilled (since $X\sim 1$ in this regime); similarly, one can check that for $\epsilon_g \ll \e \ll \Delta$ the inequality in \eref{assumpDyn} is satisfied -- in fact, it is satisfied so long as $|\e/\Delta- 1| \gg 1/(\tau_N\Delta)^2$.
However, for $|\e/\Delta-1|\lesssim 1/(\tau_N\Delta)^2$ the additional condition $\sqrt{\tau_S\Delta}\ll\tau_N\Delta$ must be met for \eref{assumpDyn} to hold; calculation of the DoS beyond this regime is outside the scope of the present work.

\subsection{Strong-coupling limit}
\label{App:Str_coup}

We now consider the case of low resistance at the $NS$ contact interface, such that at least one the two dimensionless coupling parameters $\tau_S\Delta$ and $\tau_N\Delta$ is small compared to 1. We start again from Eq.~(\ref{eq:X_full}) and make the assumption 
\be\label{sc_assump1}
|2X\tau_N\e+\tau_N^2\e^2|\ll 1 \,.
\ee
Using this assumption we simplify Eq.~(\ref{eq:X_full}) to $-\frac{\e}{\Delta}+\frac{X}{\sqrt{1+X^2}}=\frac{\tau_N\e}{\tau_S\Delta}$; solving for $X$, we find the DoS in the BCS-like form
\begin{align}
\label{eq:BCS_like_StrCoup}
n(\e)\simeq n_{>s} (\epsilon) \equiv \mathrm{Re}\left[\frac{\e}{\sqrt{\e^2-\tilde{\e}^2_\mathrm{gs}}}\right],
\end{align}
where the (approximate) minigap energy in this limit is $\tilde{\e}_\mathrm{gs}=\frac{\tau_S\Delta}{\tau_S + \tau_N}$; these results agree with those reported in Ref.~\cite{Fominov}. 

Requiring the assumption \rref{sc_assump1} to be valid as $\e \to \tilde{\e}_\mathrm{gs}$, we find the conditions
\be
\tau_N \tilde{\e}_{gs} \ll 1\, , \quad \e/\tilde{\e}_\mathrm{gs}-1 \gg (\tau_N\tilde{\e}_\mathrm{gs})^2.
\ee
The first condition can be rewritten as $1/\tau_N\Delta+ 1/\tau_S\Delta \gg 1$ and it is indeed satisfied under the assumption made at the beginning of this subsection. The second condition indicates that the BCS-like behavior is not valid close to the minigap, similar to the weak-coupling regime. Therefore, to find a more accurate position for the minigap and the behavior of the DoS near it, we take an approach similar to that of the previous subsection. Namely, let us introduce the new variable $\eta = 1/X$, and make the assumptions
\be\label{sc_assump2}
\tau_N\e \ll \eta \ll 1 \, .
\ee
Then \eref{eq:X_full} can be rewritten as $\mathcal{G}(\eta,\e) =0$ with, keeping only next to leading order terms,
\begin{align}
\label{eq:X_StrCoup_cubic}
 \mathcal{G}(\eta,\e)\equiv \eta^3-\tau_N\e \eta^2 - 2\eta \left(1-\frac{\e}{\tilde{\e}_\mathrm{gs}}\right) + 2\tau_N\e\left(1-\frac{\e}{\Delta}\right)\,.
\end{align}
Here the quadratic term can be neglected in comparison with the cubic one, see \eref{sc_assump2}. The resulting third order polynomial can be studied following the same procedure as for the weak coupling case. We then find for the minigap
\begin{align}
\label{eq:minigap_StrCoup}
\e_\mathrm{gs}\simeq\tilde{\e}_\mathrm{gs}\left[1-\frac{3}{2}(\tau_n\tilde{\e}_\mathrm{gs})^{2/3}(\frac{\tau_N}{\tau_N+\tau_S})^{2/3}\right],
\end{align}
valid when
\be
\tau_S \Delta \ll 1 \quad \mathrm{or} \quad \tau_N\Delta \ll \frac{1}{\tau_S\Delta} \lesssim 1 \, ;
\ee
this condition follows from the first inequality in \eref{sc_assump2}.

\begin{figure}[bt] 
\begin{center}
\includegraphics[width=0.48\textwidth]{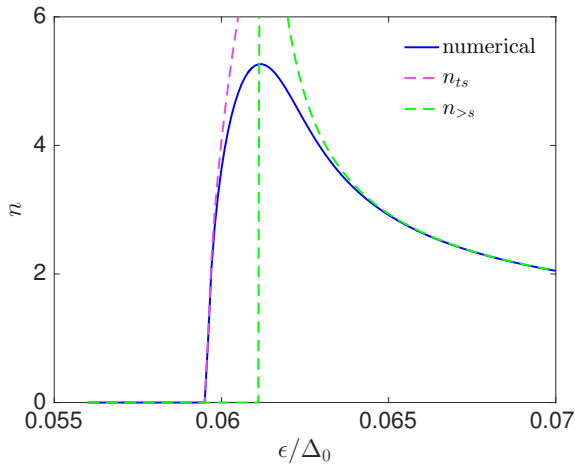}
\end{center}
\caption{(Color online) Density of states in the superconducting layer in the presence of strong proximity effect, $\tau_S\Delta_0 = 0.1$ and $\tau_N/\tau_S = 0.8$. The solid line is calculated by numerically solving the Usadel equation (\ref{eq:X_full}). Dashed lines: approximate analytical expressions just above the minigap, $n_{ts}$ of \eref{eq:StrCoup_dos}, and at higher energies, $n_{>s}$ of \eref{eq:BCS_like_StrCoup}.} \label{fig:Uniform_StrCop}
\end{figure}

\begin{figure}[t] 
\includegraphics[width=0.48\textwidth]{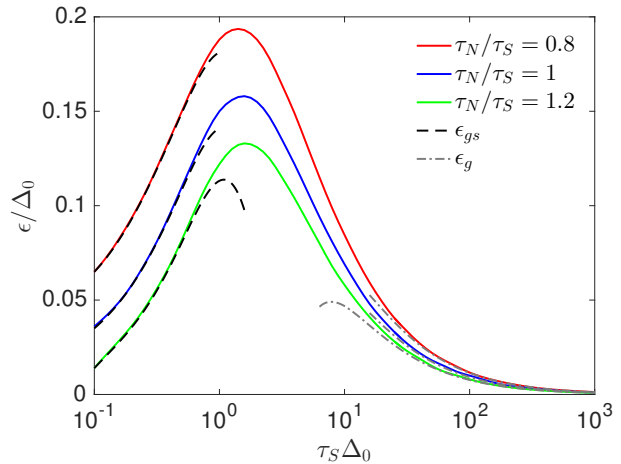}
\caption{(Color online) Normalized minigap energy $\epsilon/\Delta_0$ as a function of dimensionless parameter $\tau_S\Delta_0$. The solid lines are obtained from numerical solutions of \eref{eq:X_full} with (top to bottom) $\tau_N/\tau_S = 0.8$, $1$, $1.2$. Dashed lines: minigap for strong proximity effect, $\epsilon_{gs}$ of \eref{eq:minigap_StrCoup}. Dot-dashed lines: minigap for weak proximity effect, $\epsilon_{g}$ of \eref{eq:minigap_WeaCoup}.} \label{fig:E_g_VS_tau_S}
\end{figure}

\begin{figure}[thb]
\begin{center}
\includegraphics[width=0.48\textwidth]{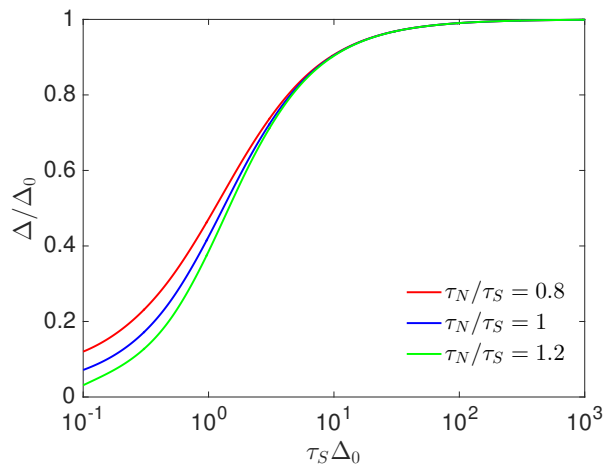}
\end{center}
\caption{(Color online) Reduction of order parameter due to proximity effect as function of the dimensionless parameter $\tau_S\Delta_{0}$ for (top to bottom) $\tau_N/\tau_S = 0.8$, $1$, $1.2$. As  $\tau_N$ increases (for example due to increased thickness of normal-metal layer) the order parameter is more strongly suppressed as the proximity effect becomes stronger (that is, $\tau_S\Delta_0$ decreases).} \label{fig:DOS_vs_tau_S}
\end{figure}

To find the density of states just above the minigap, we perform an expansion as in \eref{eq:WeakCoup_expansion_general} and finally arrive at
\begin{align}
\label{eq:StrCoup_dos}
n(\e)\simeq n_{ts}(\epsilon) \equiv \sqrt{\frac{2}{3}}\left(\frac{\tau_S+\tau_N}{\tau_N^2 \epsilon_{gs}}\right)^{2/3}
\sqrt{\frac{\e}{\e_\mathrm{gs}}-1},
\end{align}
which remains valid so long as
\begin{align}
\label{eq:StrCoup_En_cond}
\frac{\e}{\e_\mathrm{gs}}-1 \ll  \left(\frac{\tau_N^2 \epsilon_{gs}}{\tau_N+\tau_S}\right)^{2/3}.
\end{align}
In Fig.~\ref{fig:Uniform_StrCop} we show the density of states for energies near the minigap for a strongly coupled bilayer, comparing the DoS obtained from the numerical solution of the Usadel equations to our analytical findings. Similarly, in Fig.~\ref{fig:E_g_VS_tau_S} we compare numerics and analytics for the minigap energy, with coupling strength ranging from the strong regime ($\tau_S\Delta_0 = 0.1$) to the weak one ($\tau_S\Delta_0 = 10^3$). Note that in both these figures we normalize energies with respect to the bulk gap $\Delta_0$, whereas analytical expression are given in terms of the self-consistent order parameter $\Delta$; the latter is calculated numerically assuming a low temperature ($T/\Delta_0 \simeq 0.01$) and rewriting \eref{eq:op_sc} as a sum over Matsubara frequencies (see also Appendix~\ref{app:num}). For reference, we report in Fig.~\ref{fig:DOS_vs_tau_S} results of such calculations. We point out that while some of our results simply confirm those in the literature (see \textit{e.g.} Ref.~\cite{Fominov}), a number of them has not been reported before, to the best of our knowledge; we mention here for instance: the more accurate expressions for the minigap energy, \esref{eq:minigap_WeaCoup} and \rref{eq:minigap_StrCoup}, the square root threshold behavior of the DoS above the minigap, \esref{eq:WeaCoup_DOS} and \rref{eq:StrCoup_dos}, the Dynes-like DoS in \eref{eq:Dynes}, and the detailed analysis of their repspective regimes of validity.

In concluding this Appendix, we mention that the treatement presented here for the strong proximity effect may become invalid: we have used the Kuprianov-Lukichev boundary conditions, \eref{eq:app:bc}, which however are valid only in the limit of low contact transparency~\cite{lambert1,lambert2}, $\mathrm{T} \ll 1$ ; at larger transparency, more general conditions should be used, see \cite{zaitsev,lambert2}. A typical few nanometers-thick aluminum oxide insulating barrier has transparency of order $\textrm{T} \sim 10^{-5}$. For metallic films with thickness in the several tens of nanometers connected by such a barrier, we estimate $\tau_S\Delta_0 \sim 10^3$-$10^4$ if aluminum is the superconductor; threfore the present treatment is valid at most down to $\tau_S\Delta_0 \sim 0.1$ if the barrier transparency is increased while other typical parameters are kept fixed.

\section{Numerical solution of the self-consistent equation for the order parameter}
\label{app:num}

Here we briefly discuss the numerical approach we use to the calculate the order parameter in a nonuniform $NS$ bilayer as in Fig.~\ref{fig:systems}, for which the problem is effectively one-dimensional due to translational invariance in the $y$ direction. We consider a finite-size system, typically extending 10 coherence lengths on each side of the normal metal edge, $x/\xi\in[-10,10]$. We discretize the $x$ coordinate by specifying a number of mesh points $x_j$ ($j=1,\, \ldots,\, M$), with the mesh denser near the ends (to properly implement boundary conditions) and near the trap edge (where the order parameter is expected to change more rapidly). We solve the self-consistent equation iteratively (cf. Ref.~\cite{belzig2}), as we explain below, and the order parameter at points not included in the mesh is obtained by spline interpolation [except for the initial guess $\Delta^{(0)}(x)$, which is given by $\Delta_s(x)$ of \eref{eq:OP_step}]. 

Denoting with $\Delta^{(l)}$ the order parameter after $l$ iterations, we calculate $\Delta^{(l+1)}$ as follows: we numerically solve the Usadel equations \rref{eq:non_uni-usadalS} and \rref{eq:non_uni-usadalN},  with $\Delta(x) = \Delta^{(l)}(x)$, for the pairing angle $\theta^{(l)}_S(\omega_k,x)$; the solution is found directly in the Matsubara representation (\textit{i.e.}, $\epsilon \to i\omega_k$ with $\omega_k = 2\pi T (k+1/2)$, $k=0,\,1,\,2,\,\ldots$). Next, we calculate the new order parameter $\Delta^{(l+1)}$ at the mesh points using the self-consistent equation \rref{eq:op_sc}:
\begin{align}
\label{eq:SC_OP}
\frac{\Delta^{(l+1)}(x_j)}{\Do}= \frac{\sum_{k=0}^{k_\mathrm{M}} \mathrm{Re}\left[\sin \theta_\mathrm{S}^{(l)}(\omega_{k},x_j)\right]}{\sum_{k=0}^{k_\mathrm{M}} \frac{\Delta_0}{\sqrt{\Delta_0^2+\omega_k^2}}}\, .
\end{align}

We define a convergence condition as
\begin{align}
\label{eq:SC_OP_Conv}
\mathcal{C}\equiv\frac{1}{M}\sum_{j=1}^{M}\left|\frac{\Delta^{N+1}(x_j,T)-\Delta^{N}(x_j,T)}{\Delta^{N}(x_j,T)}\right|<c_0.
\end{align}
for some small number $c_0\ll 1$, and we repeat the above steps until this condition is satisfied.
In our numerical analysis, we take $T/\Delta_0=0.01$ (corresponding to about 20~mK in Al), keep $k_\mathrm{M}=2000$ Matsubara frequencies in the sums, and set $c_0=10^{-5}$. The number of iterations needed to reach convergence depends on the initial assumption; using \eref{eq:OP_step} as the starting point, convergence is usually reached within 20 iterations in the regime of weak proximity effect.

\section{Spatial evolution of single-particle density of states away from the trap}
\label{App:DOS_evolution}

Here we outline the derivation of \esref{eq:n_vs_x}-\rref{eq:p_above}, starting from the definitions for $n$ and $p$ in \esref{eq:DOS} and \rref{eq:p_def}, respectively. According to \eref{eq:Ana_theta_L}, in the uncovered section of the superconductor the pairing angle $\theta_L$ is the sum of the BCS angle $\theta_{BCS}$ and a correction. Assuming the correction to be small, $|\theta_{BCS}-\theta_{Su}| \ll \theta_{BCS}$, the corrections to $n$ and $p$ are then
\be\label{eq:dn}
\delta n(\epsilon,x) \simeq \frac12 \mathrm{Re} \left\{\sin\theta_{BCS} \left[\theta_{BCS}(\epsilon)- \theta_{Su}(\epsilon)\right]e^{\frac{x}{\xi}\sqrt{2\alpha_1(\epsilon)}}\right\}
\ee
and
\be\label{eq:dp}
\delta p(\epsilon,x) \simeq \frac12 \mathrm{Im} \left\{\cos\theta_{BCS} \left[\theta_{Su}(\epsilon)- \theta_{BCS}(\epsilon)\right]e^{\frac{x}{\xi}\sqrt{2\alpha_1(\epsilon)}}\right\}\!.
\ee

\begin{figure}[bt]
\begin{center}
\includegraphics[width=0.48\textwidth]{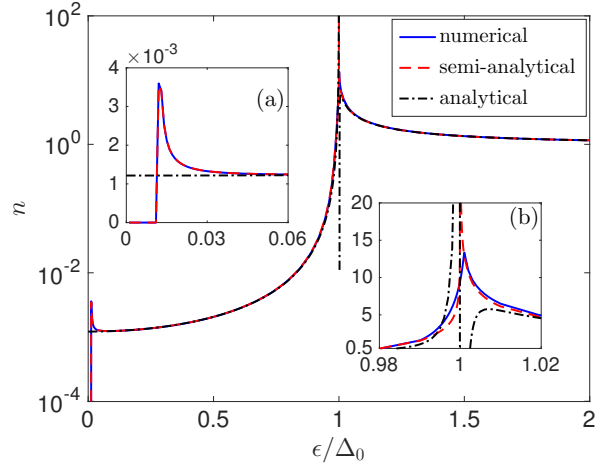}
\end{center}
\caption{(Color online) Density of states near the trap edge, $x/\xi=-1$, for $\tau_S\Delta_0 = 10^2$ and $\tau_N/\tau_S = 0.8$. Solid line (blue): self-consistent numerical approximation; dashed (red): semi-analytical approach; dot-dashed (black): analytical formulas (see text for more details). The insets (a) and (b) zoom into the minigap and gap regions, respectively.} \label{fig:DOS_Num_VS_semAn_Vs_An}
\end{figure}

At most energies (except near the gap and the minigap, see Appendix~\ref{app:unifweak}), we have $\theta_{Su} \simeq \theta_{Dy}$ of \eref{eq:thetaDynes}. Using from now on this approximation, we continue by noting the identity
\be
\tan \left(\theta_{BCS}-\theta_{Dy}\right) = i \frac{\Delta_0 \left(\epsilon+i/\tau_S\right) -\epsilon\Delta_{NS}}{\epsilon\left(\epsilon+i/\tau_S\right)-\Delta_0\Delta_{NS}}\, ,
\ee
where we used \eref{eq:thetaBCS} for $\theta_{BCS}$.
According to \eref{eq:D_NS}, at leading order we can approximate $\Delta_{NS} \simeq \Delta_0 - 1/\tau_S$, and assuming $\left|\epsilon-\Delta_0\right| \gg 1/\tau_S$ we can simplify the right hand side of the above equation and linearize its left hand side to arrive at
\be
\theta_{BCS}-\theta_{Dy} \simeq \frac{1-i}{\tau_S\Delta_0} \frac{\Delta_0^2}{\Delta_0^2-\epsilon^2}\, .
\ee
Substituting this expression into \esref{eq:dn} and \rref{eq:dp}, we find \esref{eq:n_vs_x}-\rref{eq:p_above}, where the leading BCS contributions are also included. In Fig.~\ref{fig:DOS_Num_VS_semAn_Vs_An} we show the density of states near the trap edge, $x/\xi=-1$, obtained in three ways: 1. from the numerical solution of the Usadel and self-consistent equations, \esref{eq:non_uni-usadalS}, \rref{eq:non_uni-usadalN}, and \rref{eq:op_sc}; 2. using the semi-analytical expression in \eref{eq:Ana_theta_L} in which $\theta_{Su}(\epsilon)$ is calculated numerically; 3. plotting the analytical formulas in \esref{eq:n_vs_x} and \rref{eq:n_above}. In their regions of validity, the semi-analytical and analytical results are in good agreement with the numerical findings; we stress that the decrease of the DoS calculated with the analytical formula as $\epsilon\to \Delta_0^+$ is an artifact due to the invalidity of the approximations employed when $\epsilon - \Delta_0 \lesssim 1/\tau_S$.

\section{Spectral function in the presence of a trap}
\label{App:Decay_rate}

Here we present in some detail the derivation of the formulas for the spectral functions reported in Sec.~\ref{S5}. As in that Section, we distinguish between the thermal equilibrium case and the non-equilibrium one, which accounts for the suppression of the quasiparticle density by the trap. In both cases, our starting points are \esref{eq:Sred}-\rref{eq:A}, with the appropriate expressions for the density of states $n$, pair amplitude $p$, and distribution function $f$.

\subsection{Thermal equilibrium}

We assume a thermal equilibrium distribution function, \eref{eq:feq}, at temperature $\epsilon_g \ll T\ll\omega \ll \Delta_0$ (\textit{i.e.}, ``cold'' quasiparticles). For the tunneling spectral density $S_\text{t}$, \eref{eq:S_t}, we split the integral in three integration regions: from the minigap $\epsilon_g$ to $\Delta_0 - \omega$, from $\Delta_0-\omega$ to $\Delta_0$, and from $\Delta_0$ to infinity. These three region correspond to the $bb$, $ba$, $aa$ types of transitions described in Sec.~\ref{S5}.

Let us consider first the highest energy integration region, for which we have approximately
\be
S_{aa}^\text{eq}(\omega) \simeq \frac{16 E_J}{\pi \Delta_0} \int_{\Delta_0}^{\infty}\!d\epsilon \, \frac{\epsilon\left(\epsilon +\omega\right)+\Delta_0^2}{\sqrt{\epsilon^2-\Delta_0^2}\sqrt{\left(\epsilon+\omega\right)^2-\Delta_0^2}}\, e^{-\epsilon/T}.
\ee
The approximations employed here are two: first, in the function $A$ of \eref{eq:A}, we keep only the leading terms for $n$ and $p$ in \esref{eq:n_above} and \rref{eq:p_above}; second, for the distribution functions we can neglect $f^\text{eq}(\epsilon+\omega)$ in comparison to unity and approximate $f^\text{eq}(\epsilon) \simeq e^{-\epsilon/T}$. It is then easy to check that $S_{aa}^\text{eq}$ takes the same form as in \eref{eq:S_w_x_qp} with the substitution $x_\qp \to x_\qp^\text{eq}$, thus proving \eref{eq:Saa}.

In the intermediate integration region, the above approximations for the distribution functions are still valid, and for $n(\epsilon+\omega)$ and $p(\epsilon+\omega)$ we can again just keep the leading terms in \esref{eq:n_above} and \rref{eq:p_above}. On the other hand, for $n(\epsilon)$ and $p(\epsilon)$ we must now use \esref{eq:n_vs_x} and \rref{eq:p_vs_x}, and therefore we have
\be\begin{split}
S_{ba}^\text{eq}(\omega) \simeq \frac{16 E_J}{\pi \Delta_0}\frac{1}{\tau_S\Delta_0} \int_{\Delta_0-\omega}^{\Delta_0}\!d\epsilon \, e^{-\sqrt{2}\frac{|x|}{\xi}\left(1-\frac{\epsilon^2}{\Delta_0^2}\right)^{1/4}} \\ \frac{\Delta_0^3\left(2\epsilon + \omega\right)}{\left(\Delta_0^2-\epsilon^2\right)^{3/2}\sqrt{\left(\epsilon+\omega\right)^2-\Delta_0^2}} \, e^{-\epsilon/T}.
\end{split}\ee
Here we have included a factor of 2 due to the presence of two identical traps symmetrically placed with respect to the trap, as discussed at the beginning of Sec.~\ref{S5}; since we are considering small, linear-order corrections [cf. text above \eref{eq:Ana_theta_L}], we can simply add the contributions from the two traps.
We note that formally the integral is divergent at the upper integration limit; however, this is due to the break-down of the used approximation for $n$ and $p$, which is valid for $\Delta_0 - \epsilon \gg 1/\tau_S$. Closer to the gap, both $n$ and $p$ are in fact finite (see Sec.~\ref{S4}), and we can neglect the contribution from this small integration region near the gap, as it is exponentially suppressed due to the $e^{-\epsilon/T}$ factor. Then, making the change of variables $\epsilon = \Delta_0 -\omega + \tilde\epsilon$, and neglecting corrections small in $\omega/\Delta_0$ and $T/\omega$, we find
\be\begin{split}
S_{ba}^\text{eq}(\omega) \simeq \frac{16 E_J}{\pi \Delta_0}\frac{1}{\tau_S\Delta_0} e^{-\sqrt{2}\frac{|x|}{\xi}\left(\frac{2\omega}{\Delta_0}\right)^{1/4}} 
e^{-\Delta_0/T} e^{\omega/T}
\\ \int_0\!d\tilde\epsilon \, \frac{\Delta_0^2}{2\omega^{3/2}\sqrt{\tilde\epsilon}} \, e^{-\tilde\epsilon/T}.
\end{split}\ee
Performing the integration, we finally arrive at \eref{eq:Sba}.

In the lowest integration region we use \esref{eq:n_vs_x} and \rref{eq:p_vs_x} to write
\be\begin{split}
S_{bb}^\text{eq}(\omega) \simeq \frac{16E_J}{\pi\Delta_0} \frac{1}{\left(\tau_S\Delta_0\right)^2}\!\int_{\epsilon_g}^{\Delta_0-\omega}\!\!\!\!\!\!\!d\epsilon\, f^\text{eq}(\epsilon)\,e^{-\sqrt{2}\frac{|x|}{\xi}\left(1-\frac{\epsilon^2}{\Delta_0^2}\right)^\frac14} \\
e^{-\sqrt{2}\frac{|x|}{\xi}\left[1-\frac{(\epsilon+\omega)^2}{\Delta_0^2}\right]^\frac14}\frac{\Delta_0^4\left[\epsilon\left(\epsilon+\omega\right)+\Delta_0^2\right]}{\left(\Delta_0^2-\epsilon^2\right)^{3/2}\left[\Delta_0^2-(\epsilon+\omega)^2\right]^{3/2}},
\end{split}\ee
where again we have taken into account the presence of two identical traps 
[while use of \esref{eq:n_vs_x} and \rref{eq:p_vs_x} is strictly speaking not justified near the minigap, the error thus introduced is small, see Appendix~\ref{App:DOS_evolution}].
Note that here we can still neglect $f(\epsilon+\omega)$ compared to unity, due to the assumption $\omega \gg T$, but we cannot make approximations for $f(\epsilon)$, for which we must use the full equilibrium expression, \eref{eq:feq}. Still, the distribution function forces $\epsilon$ to be small compared to $\Delta_0$, so that neglecting terms small in $T/\Delta_0$ and $\omega/\Delta_0$ we have
\be
S_{bb}^\text{eq}(\omega) \simeq \frac{16E_J}{\pi\Delta_0} \frac{1}{\left(\tau_S\Delta_0\right)^2} e^{-2\sqrt{2}\frac{|x|}{\xi}}\int_0 d\epsilon\,f^\text{eq}(\epsilon),
\ee
where we used the assumption $T\gg \epsilon_g$ to set the lower integration limit to 0. After integration, we obtain \eref{eq:Sbb}.

In contrast to quasiparticle tunneling, the pair process contribution to the spectral function, $S_\text{p}$ of \eref{eq:S_b}, is non-vanishing only it there is a finite subgap density of states (so long as $\omega < 2\Delta_0$). In fact, the integration limits in \eref{eq:S_b} are $\epsilon_g$ and $\omega-\epsilon_g$, since the function $A(\epsilon,\omega-\epsilon)$ vanishes outside this region; this further requires $\omega>2\epsilon_g$ in order for $S_\text{p}$ to be non-zero. Since $\epsilon < \omega \ll \Delta_0$, we find the approximate expression
\be
A(\epsilon,\omega-\epsilon) \simeq \frac{16E_J}{\pi\Delta_0}\frac{1}{(\tau_S\Delta_0)^2}e^{-2\sqrt{2}\frac{|x|}{\xi}},
\ee
which we obtain after substituting \esref{eq:n_vs_x} and \rref{eq:p_vs_x} into \eref{eq:A}, accounting for two traps, and neglecting terms small in $\omega/\Delta_0$. Using this expression in \eref{eq:S_b} we find
\be\label{eq:Sp_app}\begin{split}
S_\text{p}^\text{eq}(\omega) \simeq & \frac{8E_J}{\pi\Delta_0}\frac{1}{(\tau_S\Delta_0)^2}e^{-2\sqrt{2}\frac{|x|}{\xi}}\\ &\int_0^\omega\!d\epsilon\,\left[1-f^\text{eq}(\epsilon)\right]\left[1-f^\text{eq}(\omega-\epsilon)\right].
\end{split}\ee
For $T\ll \omega$, the integral in the second line gives $\omega-2T\ln2$, thus proving \eref{eq:Sp_eq}.

\begin{figure}[!b]
\begin{center}
\includegraphics[width=0.47\textwidth]{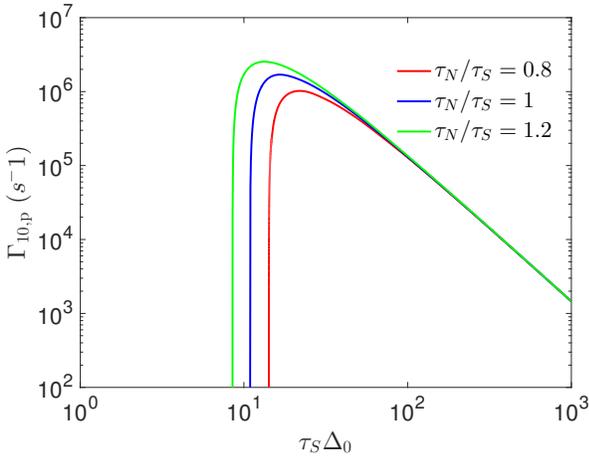}
\end{center}
\caption{(Color online) Relaxation rate $\Gamma_{10,\mathrm{p}}$ due to Cooper pair breaking. Here the temperature is set to zero and the trap is placed next to the junction, $x/\xi=0$; other parameters are as in Fig.~\ref{fig:rates_TherEqui}. 
The fast vanishing of the relaxation rate at $\tau_S\Delta_0 \sim 10$ is due to the violation of condition in \eref{eq:app:condition_S_w_b}.}
\label{fig:breaking_vs_tau_S}
\end{figure}

We have remarked above that the condition 
\be\label{eq:app:condition_S_w_b}
\omega_{10}>2\epsilon_g
\ee
is necessary for $S_\text{p}$ to be finite, and we have shown
in Fig.~\ref{fig:E_g_VS_tau_S} that the magnitude of the minigap energy has a non-monotonic behaviour with respect to the parameter $\tau_S\Delta_0$ whose inverse quantifies the strength of the proximity effect. Indeed, as $\tau_S\Delta_0$ decreases, the spectral function $S_\text{p}$ and hence its contribution to the relaxation rate increase as $1/(\tau_S\Delta_0)^2$. However, because of the increase in the minigap with decreasing $\tau_S\Delta_0$, the condition in \eref{eq:app:condition_S_w_b} can potentially be violated; in this case qubit relaxation due to pair processes is no longer possible. This is evident in Fig.~\ref{fig:breaking_vs_tau_S}, which shows the relaxation rate $\Gamma_{10,\mathrm{p}}$ due to Cooper-pair breaking as a function of $\tau_S\Delta_0$; the rate is calculated by setting $T=0$, in which case the spectral function is
\be
S_\text{p}^\text{eq}(\omega) \simeq \frac{8E_J}{\pi\Delta_0}\frac{1}{(\tau_S\Delta_0)^2}e^{-2\sqrt{2}\frac{|x|}{\xi}}\left(\omega-2\epsilon_g\right)\theta\left(\omega-2\epsilon_g\right),
\ee
and using this expression in \eref{eq:g10_1}. We note that if $\tau_S\Delta_0$ is decreased further, the decay rate will become finite again (cf. Fig.~\ref{fig:E_g_VS_tau_S}), but for such small values of $\tau_S\Delta_0$ \eref{eq:Sp_eq} is not applicable, since it was derived under the assumption $\tau_S\Delta_0 \gg 1$.

\subsection{Suppressed quasiparticle density}
\label{app:supp}

We now consider the case in which we account for the suppression of the quasiparticle density by the trap using the distribution function of the form in \eref{eq:fnoneq}, with the position-dependent chemical potential $\tilde\mu$ given by \eref{eq:mu}.
As in the previous subsection, we assume two identical, symmetrically placed traps (without writing this assumption explicitly anymore for brevity) and $\epsilon_g\ll\tilde{T}\ll\omega\ll\Delta_0$, and to evaluate the tunneling spectral density $S_\text{t}$ we again split the integration region into three intervals: $[\epsilon_g,\Delta_0-\omega]$, $[\Delta_0-\omega,\Delta_0]$, and $[\Delta_0,\infty]$, leading to the contributions $\tilde{S}_{bb}$, $\tilde{S}_{ba}$, and $\tilde{S}_{aa}$, respectively. For $\tilde{S}_{aa}$ and $\tilde{S}_{ba}$ we can perform the calculation as described above for the thermal equilibrium case, although additional constraints are needed to justify the approximations that involve the additional parameter $\tilde\mu$: for $\tilde{S}_{aa}$ we need $(\Delta_0-\tilde\mu)/\tilde{T} \gtrsim 1$, while for $\tilde{S}_{ba}$ the somewhat more restrictive condition $\exp\left[\left(\Delta_0 -\tilde\mu -\omega\right)/\tilde{T}\right]\gg 1$ is required. When these conditions are met, we find that the spectral densities have the same form as in thermal equilibrium, see \esref{eq:tSaa} and \rref{eq:tSba}.

The calculation of $\tilde{S}_{bb}$ is more involved, as different parameter regimes must be distinguished. The simplest case is that of $\tilde\mu\lesssim \omega$; then we can proceed as in the above derivation of $S_{bb}^\text{eq}$, the only difference being that we have to explicitly keep both $f(\epsilon)$ and $f(\epsilon+\omega)$, without approximations. This way we find that the expression for $\tilde{S}_{bb}$ is obtained by the replacement $T\to \tilde{T} \ln[(1+e^{\tilde\mu/\tilde{T}})/(1+e^{(\tilde\mu-\omega)/\tilde{T}})]/\ln2$ in the formula \eref{eq:Sbb} for $S_{bb}^\text{eq}$ (note that the replacement simplifies to $T\to\tilde{T}$ for $\tilde\mu\ll\tilde{T}$, as could be expected). When $\tilde\mu\gg\omega$ (which in practice means that $\tilde\mu$ is comparable to, albeit smaller than, $\Delta_0$), the approximate calculation of $\tilde{S}_{bb}$ is different: because of the combination of distribution functions, the main contribution to the integral comes from the region between $\tilde\mu-\omega$ and $\tilde\mu$. In fact, for $\tilde{T} \to 0$ the combination reduces to $\theta(\epsilon-\tilde\mu+\omega)\theta(\tilde\mu-\epsilon)$, and this form gives the leading contribution in the low-temperature regime so long as $(\Delta_0 -\tilde\mu-\omega)/\tilde{T} \gg 1$. Using this step function approximation for the distribution functions, in the resulting finite integration region almost all the other factors in the integral are approximately constant, except for $(\Delta_0^2-(\epsilon+\omega)^2)^{-3/2} \simeq (\Delta_0 +\tilde\mu)^{-3/2}(\Delta_0 -\epsilon-
\omega)^{-3/2}$. Integrating the last factor, we finally arrive at 
\begin{align}
& \tilde{S}_{bb} \simeq \frac{16E_J}{\pi\Delta_0}\frac{1}{\left(\tau_S\Delta_0\right)^2}e^{-2\sqrt{2}\frac{|x|}{\xi}\left(1-\frac{\tilde\mu^2}{\Delta_0^2}\right)^\frac14} \\ & \frac{2\Delta_0^4\left(\Delta_0^2+\tilde\mu^2\right)}{\left(\Delta_0^2-\tilde\mu^2\right)^{3/2}(\Delta_0+\tilde\mu)^{3/2}} 
 \left(\frac{1}{\sqrt{\Delta_0 - \tilde\mu-\omega}} - \frac{1}{\sqrt{\Delta_0 -\tilde\mu}} \right)\!.
\nonumber
\end{align}
The results for the two regimes are summarized in \eref{eq:tSbb}.

Turning our attention to pair processes, the calculation of $\tilde{S}_\text{p}$ is once again similar to that in thermal equilibrium: we can simply replace $f^\text{eq}$ with $f$ of \eref{eq:fnoneq} in \eref{eq:Sp_app}. After integration over $\epsilon$ we find the expression given in \eref{eq:tSp}.



\begin{thebibliography}{30}
\bibitem{Day}P. K. Day, H. G. LeDuc, B. A. Mazin, A. Vayonakis, and J. Zmuidzinas, \href{https://doi.org/10.1038/nature02037}{Nature \textbf{425}, 817 (2003).}
\bibitem{PekolaRMP}J. P. Pekola, O.-P. Saira, V. F. Maisi, A. Kemppinen, M. M\"ott\"onen, Y. A. Pashkin, and D. V. Averin, \href{https://doi.org/10.1103/RevModPhys.85.1421}{Rev. Mod. Phys. \textbf{85}, 1421 (2013).}
\bibitem{GC_PRL106}G. Catelani, J. Koch, L. Frunzio, R. J. Schoelkopf, M. H. Devoret, and L. I. Glazman, \href{https://doi.org/10.1103/PhysRevLett.106.077002}{Phys. Rev. Lett. \textbf{106}, 077002 (2011).}
\bibitem{Paik}H. Paik, D. I. Schuster, L. S. Bishop, G. Kirchmair, G. Catelani, A. P. Sears, B. R. Johnson, M. J. Reagor,
L. Frunzio, L. I. Glazman, S. M. Girvin, M. H. Devoret, and R. J. Schoelkopf, \href{https://journals.aps.org/prl/abstract/10.1103/PhysRevLett.107.240501}{Phys. Rev. Lett. \textbf{107}, 240501 (2011).}
\bibitem{Visser}P. J. de Visser, J. J. A. Baselmans, P. Diener, S. J. C. Yates, A. Endo, and T. M. Klapwijk, \href{https://doi.org/10.1103/PhysRevLett.106.167004}{Phys. Rev. Lett. \textbf{106}, 167004 (2011).}  
\bibitem{Riste}D. Rist\`e, C. C. Bultink, M. J. Tiggelman, R. N. Schouten, K. W. Lehnert, and L. DiCarlo, \href{10.1038/ncomms2936}{Nat. Commun. \textbf{4}, 1913 (2013).}
\bibitem{Aumentado}J. Aumentado, M. W. Keller, J. M. Martinis, and M. H. Devoret, \href{https://doi.org/10.1103/PhysRevLett.92.066802}{Phys. Rev. Lett. \textbf{92}, 066802 (2004).}
\bibitem{Sun}L. Sun, L. DiCarlo, M. D. Reed, G. Catelani, L. S. Bishop, D. I. Schuster, B. R. Johnson, Ge A. Yang,
L. Frunzio, L. Glazman, M. H. Devoret, and R. J. Schoelkopf, \href{https://journals.aps.org/prl/abstract/10.1103/PhysRevLett.108.230509}{Phys. Rev. Lett. \textbf{108}, 230509 (2012).}
\bibitem{Nsanzineza}I. Nsanzineza and B. L. T. Plourde, \href{https://doi.org/10.1103/PhysRevLett.113.117002}{Phys. Rev. Lett. \textbf{113}, 117002 (2014).}
\bibitem{Wang}C. Wang, Y.Y. Gao, I.M. Pop, U. Vool, C. Axline, T. Brecht, R.W. Heeres, L. Frunzio, M.H. Devoret,
G. Catelani, L.I. Glazman, and R.J. Schoelkopf, \href{http://www.nature.com/articles/ncomms6836}{ Nat. Commun. \textbf{5}, 5836 (2014).}
\bibitem{Taupin}M. Taupin, I. M. Khaymovich, M. Meschke, A. S. Melnikov, and J. P. Pekola, \href{http://www.nature.com/articles/ncomms10977}{Nat. Commun. \textbf{7},  10977 (2016).}
\bibitem{Knowles}H. S. Knowles, V. F. Maisi, and J. P. Pekola, \href{http://aip.scitation.org/doi/pdf/10.1063/1.4730407}{ Appl. Phys. Lett.\textbf{100}, 262601 (2012).}
\bibitem{Pekola2000}J. P. Pekola, D. V. Anghel, T. I. Suppula, J. K. Suoknuuti, A. J. Manninen, and M. Manninen, \href{http://aip.scitation.org/doi/pdf/10.1063/1.126474}{Appl. Phys. Lett. \textbf{76}, 2782 (2000).}
\bibitem{Rajauria}S. Rajauria, H. Courtois, and B. Pannetier, \href{https://journals.aps.org/prb/abstract/10.1103/PhysRevB.80.214521}{Phys. Rev. B \textbf{80}, 214521 (2009).}
\bibitem{Patel}U. Patel, I. V. Pechenezhskiy, B. L. T. Plourde, M. G. Vavilov, R. McDermott,
\href{https://doi.org/10.1103/PhysRevB.96.220501}{Phys. Rev. B \textbf{96}, 220501(R) (2017).}
\bibitem{Riwar}R.-P. Riwar, A. Hosseinkhani, L. D. Burkhart, Y. Y. Gao, R. J. Schoelkopf, L. I. Glazman, and G. Catelani, \href{https://journals.aps.org/prb/abstract/10.1103/PhysRevB.94.104516}{Phys. Rev. B \textbf{94}, 104516 (2016).}
\bibitem{Gennes1}P. G. de Gennes and E. Guyon, \href{https://doi.org/10.1016/0031-9163(63)90401-3}{Phys. Lett. \textbf{3}, 168 (1963).}
\bibitem{Gennes2}P. G. de Gennes, \href{https://journals.aps.org/rmp/abstract/10.1103/RevModPhys.36.225}{Rev. Mod. Phys. \textbf{36}, 225 (1964).}
\bibitem{McMilan}W. L. McMillan, \href{https://journals.aps.org/pr/abstract/10.1103/PhysRev.175.537}{Phys. Rev. \textbf{175}, 537 (1968).}
\bibitem{Gueron}S. Gueron, H. Pothier, N. O. Birge, D. Esteve, and M. H. Devoret, \href{https://journals.aps.org/prl/abstract/10.1103/PhysRevLett.77.3025}{Phys. Rev. Lett. \textbf{77}, 3025 (1996).}
\bibitem{belzig}W. Belzig, C. Bruder, and G. Sch\"on, \href{https://doi.org/10.1103/PhysRevB.54.9443}{Phys. Rev. B \textbf{54}, 9443 (1996).}
\bibitem{Moussy}N. Moussy, H. Courtois and B. Pannetier, \href{https://doi.org/10.1209/epl/i2001-00361-2}{Europhys. Lett. \textbf{55}, 861 (2001).}
\bibitem{kup}A. A. Golubov and M. Yu. Kupriyanov, \href{https://doi.org/10.1007/BF00683247}{J.  Low Temp. Phys. \textbf{70}, 83 (1988).}
\bibitem{Zhou}F. Zhou, P. Charlat, B. Spivak, B. Pannetier, \href{https://doi.org/10.1023/A:1022628927203}{J.  Low Temp. Phys. \textbf{110}, 841 (1998).}
\bibitem{Ivanov}D. A. Ivanov, R. von Roten, and G. Blatter, \href{https://doi.org/10.1103/PhysRevB.66.052507}{Phys. Rev. B \textbf{66}, 052507 (2002).}
\bibitem{Hammer}J. C. Hammer, J. C. Cuevas, F. S. Bergeret, and W. Belzig, \href{https://doi.org/10.1103/PhysRevB.76.064514}{Phys. Rev. B \textbf{76}, 064514 (2007).}
\bibitem{Sueur}H. le Sueur, P. Joyez, H. Pothier, C. Urbina, and D. Esteve, \href{https://doi.org/10.1103/PhysRevLett.100.197002}{Phys. Rev. Lett. \textbf{100}, 197002 (2008).} 
\bibitem{Kauppila}V. J. Kauppila, H. Q. Nguyen, and T. T. Heikkil\"a, \href{https://doi.org/10.1103/PhysRevB.88.075428}{Phys. Rev. B \textbf{88}, 075428 (2013).}
\bibitem{Cherkez}V. Cherkez, J. C. Cuevas, C. Brun, T. Cren, G. M\`enard, F. Debontridder, V. S. Stolyarov, and D. Roditchev, \href{https://doi.org/10.1103/PhysRevX.4.011033}{Phys. Rev. X \textbf{4}, 011033 (2014).}
\bibitem{Fominov}Ya. V. Fominov and M. V. Feigelman, \href{http://journals.aps.org/prb/pdf/10.1103/PhysRevB.63.094518}{Phys. Rev. B \textbf{63}, 094518 (2001).}
\bibitem{Leppakangas}J. Lepp\"akangas and M. Marthaler, \href{https://journals.aps.org/prb/abstract/10.1103/PhysRevB.85.144503}{Phys. Rev. B \textbf{85}, 144503 (2012).}
\bibitem{GC_PRB11}G. Catelani, R. J. Schoelkopf, M. H. Devoret, and L. I. Glazman, \href{https://journals.aps.org/prb/abstract/10.1103/PhysRevB.84.064517}{Phys. Rev. B \textbf{84}, 064517 (2011).}
\bibitem{Hosseinkhani}A. Hosseinkhani, R.-P. Riwar, R. J. Schoelkopf, L. I. Glazman, and G. Catelani, Phys. Rev. Applied, accepted
(\href{https://arxiv.org/abs/1706.09336}{arXiv:1706.09336}).
\bibitem{Rammer}J. Rammer and H. Smith, \href{https://doi.org/10.1103/RevModPhys.58.323}{Rev. Mod. Phys. \textbf{58}, 323 (1986).} 
\bibitem{Chandrasekhar}V. Chandrasekhar, in \textit{The Physics of Superconductors}, K.-H. Bennemann and J. B. Ketterson Eds. (Springer-Verlag, Berlin, 2004), Vol. II.
\bibitem{belzig2}W. Belzig, F. K. Wilhelm, C. Bruder, G. Sch\"on, A. D. Zaikin, \href{https://doi.org/10.1006/spmi.1999.0710}{Superlatt. Microstruct \textbf{25}, 1251 (1999).}
\bibitem{Kopnin}N. Kopnin, \textit{Theory of Nonequilibrium Superconductivity} (Oxford University Press, New York, 2001).
\bibitem{Usadel}K. D. Usadel, \href{https://doi.org/10.1103/PhysRevLett.25.507}{Phys. Rev. Lett. \textbf{25}, 507 (1970).} 
\bibitem{Dynes}R. C. Dynes, V. Narayanamurti, and J. P. Garno, \href{http://journals.aps.org/prl/abstract/10.1103/PhysRevLett.41.1509}
{Phys. Rev. Lett. \textbf{41}, 1509 (1978).}
\bibitem{Owen}C. S. Owen and D. J. Scalapino, \href{https://journals.aps.org/prl/pdf/10.1103/PhysRevLett.28.1559}{Phys. Rev. Lett. \textbf{28}, 1559 (1972).}
\bibitem{Scalapino}S. B. Kaplan, C. C. Chi, D. N. Langenberg, J. J. Chang, S. Jafarey, and D. J. Scalapino, \href{https://journals.aps.org/prb/abstract/10.1103/PhysRevB.14.4854}{Phys. Rev. B \textbf{14}, 4854 (1976).}
\bibitem{Ansari}M. H. Ansari, \href{http://iopscience.iop.org/article/10.1088/0953-2048/28/4/045005/meta}{Supercond. Sci. Technol. \textbf{28}, 045005 (2015).}
\bibitem{Vool}U. Vool, I. M. Pop, K. Sliwa, B. Abdo, C. Wang, T. Brecht, Y. Y. Gao, S. Shankar, M. Hatridge, G. Catelani, M. Mirrahimi, L. Frunzio, R. J. Schoelkopf, L. I. Glazman, and M. H. Devoret, \href{https://journals.aps.org/prl/abstract/10.1103/PhysRevLett.113.247001}{Phys. Rev. Lett. \textbf{113}, 247001 (2014).}
\bibitem{Jin}X. Y. Jin, A. Kamal, A. P. Sears, T. Gudmundsen, D. Hover, J. Miloshi, R. Slattery, F. Yan, J. Yoder, T. P. Orlando, S. Gustavsson, and W. D. Oliver, \href{https://journals.aps.org/prl/abstract/10.1103/PhysRevLett.114.240501}{Phys. Rev. Lett. \textbf{114}, 240501 (2015).}
\bibitem{loss}D. Rainis and D. Loss, \href{https://doi.org/10.1103/PhysRevB.85.174533}{Phys. Rev. B \textbf{85}, 174533 (2012).}
\bibitem{Rod}L. Serrier-Garcia, J. C. Cuevas, T. Cren, C. Brun, V. Cherkez, F. Debontridder, D. Fokin, F. S. Bergeret, and D. Roditchev, \href{https://doi.org/10.1103/PhysRevLett.110.157003}{Phys. Rev. Lett. \textbf{110}, 157003 (2013).}
\bibitem{Lukichev}M. Yu. Kupriyanov and V. F. Lukichev, \href{http://www.jetp.ac.ru/cgi-bin/e/index/e/67/6/p1163?a=list}{Sov. Phys. JETP \textbf{67}, 1163 (1988).}
\bibitem{lambert1}C. J. Lambert, R. Raimondi, V. Sweeney, and A. F. Volkov, \href{https://doi.org/10.1103/PhysRevB.55.6015}{Phys. Rev. B \textbf{55}, 6015 (1997).}
\bibitem{lambert2}C. J. Lambert and R. Raimondi, \href{https://doi.org/10.1088/0953-8984/10/5/003}{J. Phys.: Condens. Matter \textbf{10}, 901 (1998).}
\bibitem{zaitsev}A. V. Zaitsev, \href{http://www.jetp.ac.ru/cgi-bin/e/index/e/59/5/p1015?a=list}{Sov. Phys. JETP \textbf{59}, 1015 (1984).}
\end{thebibliography}
\end{document}